\documentclass[aps,pra,twocolumn,letterpaper,superscriptaddress,10pt]{revtex4-1}
\usepackage{amssymb,amsthm,amsmath,amsfonts}
\usepackage{graphicx,ulem,mathptmx,textcomp}
\usepackage[pdftex,dvipsnames,usenames]{xcolor}
\usepackage[colorlinks=true,urlcolor=blue,citecolor=blue,linkcolor=blue]{hyperref}

\begin{document}

\title{Quantum photonics with active feedback loops}

\author{M. Engelkemeier}
	\affiliation{Integrated Quantum Optics Group, Applied Physics, University of Paderborn, 33098 Paderborn, Germany}

\author{L. Lorz}
	\affiliation{Integrated Quantum Optics Group, Applied Physics, University of Paderborn, 33098 Paderborn, Germany}

\author{Syamsundar De}
	\affiliation{Integrated Quantum Optics Group, Applied Physics, University of Paderborn, 33098 Paderborn, Germany}

\author{B. Brecht}
	\affiliation{Integrated Quantum Optics Group, Applied Physics, University of Paderborn, 33098 Paderborn, Germany}

\author{I. Dhand}
	\affiliation{Institut f\"ur Theoretische Physik and Center for Integrated Quantum Science and Technology (IQST), University of Ulm, 89069 Ulm, Germany}
	\affiliation{Xanadu Quantum Technologies, 777 Bay Street, Toronto ON, M5G 2C8, Canada}

\author{M. B. Plenio}
	\affiliation{Institut f\"ur Theoretische Physik and Center for Integrated Quantum Science and Technology (IQST), University of Ulm, 89069 Ulm, Germany}

\author{C. Silberhorn}
	\affiliation{Integrated Quantum Optics Group, Applied Physics, University of Paderborn, 33098 Paderborn, Germany}

\author{J. Sperling}
	\email{jan.sperling@upb.de}
	\affiliation{Integrated Quantum Optics Group, Applied Physics, University of Paderborn, 33098 Paderborn, Germany}

\date{\today}

\begin{abstract}
	We develop a unified theoretical framework for the efficient description of multiphoton states generated and propagating in loop-based optical networks which contain nonlinear elements.
	These active optical components are modeled as nonlinear media, resembling a two-mode squeezer.
	First, such nonlinear components can be seeded to manipulate quantum states of light, as such enabling photon addition protocols.
	And, second, they can function as an amplifying medium for quantum light.
	To prove the practical importance of our approach, the impact of multiple round trips is analyzed for states propagating in experimentally relevant loop configurations of networks, such as time-multiplexed driven quantum walks and iterative photon-number state generation protocols.
	Our method not only enables us to model such complex systems but also allows us to propose alternative setups that overcome previous limitations.
	To characterize the systems under study, we provide exact expressions for fidelities with target states, success probabilities of heralding-type measurements, and correlations between optical modes, including many realistic imperfections.
	Moreover, we provide an easily implementable numerical approach by devising a vector-type representation of photonic states, measurement operators, and passive and active processes.
\end{abstract}

\maketitle

\section{Introduction}

	In classical communication technologies, nonlinear elements play a crucial role when it comes to routing, amplifying, and, in general, manipulating light traveling in optical fibers over long distances.
	Similarly, feedback loops allow for active and dynamical responses of a network's node to incident light and control signals for carrying out processing tasks.
	Beyond classical applications, quantum communication protocols are on the verge of becoming a practical means of sending and receiving sensitive information \cite{GT07,KMNRDM07,S09}, e.g., via quantum key distribution \cite{DYDSS08,TNZHHTY07}.
	However, certain quantum laws, such as the no-cloning theorem \cite{P70,WZ82}, set fundamental limitations on the operation of quantum communication nodes.
	Likewise, the uncertainty principle provides a lower bound to the excess noise that is unavoidable when amplifying quantum signals \cite{KOR95,JSCLA06}.
	Thus, a rigorous analysis in the quantum domain is paramount for an advantageous utilization of active elements and feedback loops in future applications.
	Also, the robustness of quantum characteristics of light under realistic conditions needs to be studied for making reliable predictions and before setting up costly experiments.

	One way to manipulate a quantum system on demand is feedback control, being a well-established tool in classical systems.
	Its extension into the quantum domain shows equally exciting promises for efficient alterations of quantum systems that are central for future quantum information technologies;
	see, e.g., Refs. \cite{WM09,ZLWJN17} for thorough overviews.
	In general, the combination of techniques from classical photonics and quantum optics defines the field of quantum photonics \cite{PMTB09,SBBT16,QZWWLGKMSRWBTM18}.
	Controlling a quantum system using feedback is, in general, classified into two categories, measurement-based \cite{W94} and quantum coherent feedback control \cite{L00}, addressing feedback procedures for measurements and states, respectively.
	Such studies also concern the control of quantum systems through nonlinear optical interaction using feedback.
	For example, time-delayed coherent quantum feedback can be used for establishing sophisticated control mechanisms \cite{G15}.

	In quantum physics in general, and in quantum optics in particular, the three components which are essential for a full quantum model are the preparation, propagation, and detection of light.
	Each of these instances comes with its own challenges and benefits when compared with a classical description of light; see Refs. \cite{MW95,VW06,A12} for detailed introductions.
	In turn, a sweet spot for the joint operation of all elements in a setup has to be determined to maximize the potential gain through the sensible usage of quantum resources.
	This demands a well-adjusted formalism to apply the underlying theory.

	The first key component of a quantum-optical system are sources of quantum light which mainly rely on nonlinear interactions of light and matter.
	Since the photon carries and distributes quantum information, the generation of single- and multiphoton states has attracted a considerable amount of attention \cite{EMP11,CZD19}.
	For instance, quantum dots offer a high-quality source of single photons \cite{LD98,MKBSPZHI00,Detal16,Wetal19}.
	Another prominent way to produce photons is the heralded, i.e., nondeterministic, generation of photons from a parametric down-conversion (PDC) process \cite{BW70,GRA86,GM87}.
	Remarkably, these technologies can produce quantum states of light which are compatible with existing optical telecommunication networks \cite{CZD19}, thus inherently combining quantum properties with an existing infrastructure.
	However, a nonunit purity of the heralded photon states can severely diminish quantum characteristics \cite{TBHLNGS19}, thus affecting their usefulness for quantum tasks.

	Second, the detection of photonic states of light, including the assessment of their quantum features, is contingent on the capability to resolve individual photons \cite{EMP11,S07,H09}.
	However, true photon-number resolution can only be approximated with currently available technologies \cite{SVA12,JB19}.
	For instance, state-of-the-art transition-edge sensors can only discern a few photons \cite{LMN08}, require well-controlled conditions \cite{I95}, and exhibit a nonlinear response to the number of photons \cite{SECMRKNLGWAV17}.
	A more practicable approach employs multiplexed detection schemes \cite{PTKJ96} in which an incident signal is split into multiple signals with reduced intensity, and each output signal is then measured with single on-off detector.
	A resource-efficient implementation of such a scheme are fiber-loop detection layouts \cite{BW03}.
	Like for transition-edge sensors, saturation effects are an example of imperfections which cap the performance of such detection devices.

	Finally, and maybe most importantly, the manipulation of light enables us to distribute quantum properties of photons over multiple parties; see, e.g., Refs. \cite{KLM01,SP19}.
	For example, passive optical networks, consisting of beam splitters and phase shifters, render it possible to convert single-mode nonclassicality into two- or multimode entanglement \cite{X02,KSBK02,WEP03,VS14}, a key resource for many quantum protocols \cite{NC00,PV07,HHHH09}.
	A fundamental application of passive optical networks are multiphoton interference experiments \cite{KBMW98,MTSKBHJBKGSLSW13,NWXC18,JC18}, generalizing the Hong-Ou-Mandel two-photon quantum phenomenon \cite{HOM87}.
	Modern applications also lie in the field of certifying quantum enhancements, e.g., through boson sampling \cite{AA13}.
	Nevertheless, both mentioned and highly relevant examples employ only static optical networks.

	Consequently, controlled quantum state manipulations via nonlinear optical elements can be expected to enlarge the family of potential quantum applications even further \cite{YWCG08,BC16}.
	Typically, those processes are driven by a pump, thus offering an active control.
	For instance, boson sampling can be generalized to driven boson sampling by using optical squeezers as a second-order nonlinear component \cite{BBSKHJS17}.
	Also, photon-addition protocols can be used to build up non-Gaussian quantum states, again relying on second-order nonlinearities as well as conditional measurements \cite{AT92,ZVB04,SVA14}.
	Note that nonlinear (specifically, non-Gaussian) processes are required for universal quantum information processing \cite{ESP02}.
	As mentioned before, nonlinear processes also impose fundamental limitations \cite{KOR95,JSCLA06} (e.g., introducing excess noise to a state), hindering an unrestricted usage of active elements to improve quantum technologies.
	Therefore, a toolbox is required that is able to unveil benefits of experiments, even under realistic conditions, to truly exploit the potential of nonlinear optics and feedback architectures in a quantum setting.

	In this article, we develop such a sought-after framework that enables us to theoretically model and devise loop-based setups which contain active elements.
	This approach not only combines nonlinear elements, actively controlled by a pump field, with feedback networks but also allows us to study different imperfections either separately or jointly, such as losses, noise, saturation effects, etc.
	Our method further enables us to comprehensively analyze the evolution of quantum features in such scenarios.
	It also leads to a closed description of a broad and practically relevant class of quantum states, quantum measurements, and quantum processes.
	Moreover, by applying our technique to state-of-the-art implementations, we are additionally able to propose schemes which favorably alter the function of existing experiments.
	To demonstrate this, we consider a sequential heralding of multiphoton states, which is achieved through a feedback mechanism, and the quantum amplification of quantum correlated light which is attenuated as it propagates in a lossy interferometer loop.
	In both cases, we show that our careful characterization results in a usage of nonlinear elements and loop configurations which can indeed improve quantum-optical properties.
	This demonstrates the unique capabilities of our approach to accurately model and further advance quantum photonics in sophisticated setups under realistic conditions.

	The paper is structured as follows:
	In Sec. \ref{sec:Prelim}, we reintroduce an apparently simple operator that, however, defines the fundamental building block for our general treatment.
	A second-order nonlinear process is exactly analyzed in Sec. \ref{sec:ActiveElement}, using an exponential-operator-based algebra, and including quantum seeds to this process and additional conditional measurements.
	In Sec. \ref{sec:Representation}, the method is generalized to a unified vector-type decomposition for photonic quantum states, measurements, and passive and active processes, including many imperfections, which is readily accessible as a numerical toolbox for our operator algebra.
	As examples of our general treatment, this methodology is then applied to setups to produce higher-order photon states, Sec. \ref{sec:Melanie}, and to amplify quantum correlations that are attenuated by loss, Sec. \ref{sec:Lennart}.
	Finally, we conclude and discuss our findings in Sec. \ref{sec:Conclusion}.

\section{Preliminaries}\label{sec:Prelim}

	For the purpose of our following studies, we consider a family of operators which are rather useful when formulating our general methodology.
	This essential element is an exponential of the photon-number operator $\hat n=\hat a^\dag\hat a$, where $\hat a$ is the annihilation operator of the quantized radiation mode under study.
	This operator takes the form
	\begin{equation}
		\label{eq:MainOperator}
		\hat E(x)=x^{\hat n}={:}\exp\left([x-1]\hat n\right){:},
	\end{equation}
	where ``${:}\bullet{:}$'' denotes the normal ordering prescription \cite{VW06}.
	For $x=1$, $x=0$, and $x=-1$, we get the identity $\hat E(1)=\hat 1$, the vacuum projector $\hat E(0)=|0\rangle\langle0|$, and the parity operator $\hat E(-1)=(-1)^{\hat n}$, respectively.
	Here, it is sufficient to restrict ourselves to values $0\leq x\leq 1$.

	Photon-number states $|n\rangle$, where $n\in\mathbb N$, can be conveniently represented through this operator via derivatives,
	\begin{equation}
		\label{eq:Photons}
		|n\rangle\langle n|={:}\frac{\hat n^n}{n!}e^{-\hat n}{:}=\frac{1}{n!}\left.\partial^n_{x}\hat E(x)\right|_{x=0},
	\end{equation}
	which, as we would like to remark, is related to the method of generating functions.
	Likewise, the above relation can be expressed in terms of the photon-number expansion
	\begin{equation}
		\label{eq:ExpExpansion}
		\hat E(x)=\sum_{n\in\mathbb N} x^n |n\rangle\langle n|.
	\end{equation}
	The above relations can be interpreted as follows: $\hat E(x)$, as a function of $x$, carries the information about all photon-number states simultaneously.
	It is additionally convenient to formulate two simple rules for a calculus that involves the exponential operators of the form \eqref{eq:MainOperator}.
	Namely, the trace over $\hat E(x)$ reads $\mathrm{tr}[\hat E(x)]=(1-x)^{-1}$, and the product of two exponential operators obeys $\hat E(x)\hat E(y) = \hat E(xy)$.

	Examples in which the operator $\hat E(x)$ is of great interest---beyond the method introduced later in this work---are the description of thermal states and on-off detectors; see, e.g., Ref. \cite{SVA14} for a comprehensive analysis.
	Specifically, the operator in Eq. \eqref{eq:MainOperator} is related to thermal states via
	\begin{equation}
		\label{eq:Thermal}
		\hat\rho_\mathrm{th}
		={:}\frac{e^{-\hat n/(\bar n+1)}}{\bar n+1}{:}
		=\frac{1}{\bar n+1}\hat E\left(\frac{\bar n}{\bar n+1}\right),
	\end{equation}
	for a mean thermal photon number $\bar n$.
	In addition, the positive operator-valued measure of an on-off detector reads
	\begin{equation}
		\label{eq:OnOff}
		\hat \Pi_\mathrm{off}={:}e^{-(\eta\hat n+\delta)}{:}=e^{-\delta}\hat E(1-\eta)
		\text{ and }
		\hat \Pi_\mathrm{on}=\hat 1-\hat \Pi_\mathrm{off},
	\end{equation}
	where $\eta$ is the quantum efficiency (likewise, $1-\eta$ defines the loss) and $\delta$ is the dark count contribution.

	In the following, we develop a technique which extensively employs operators of the form \eqref{eq:MainOperator}.
	In fact, every component of our treatment, including states, processes, and measurements, can be expressed via linear combinations and mappings of the simple operator $\hat E(x)$.

\section{Single-pass nonlinear process}\label{sec:ActiveElement}

	As another key ingredient of our treatment to analyze complex experimental settings, we describe the action of an active medium when a quantum state passes it once.
	The specific process under study resembles a second-order two-mode nonlinear optical process, controlled by a pump field \cite{MG67b,MG67a}.

\subsection{Two-mode squeezing transformations}

	The nonlinear medium in our setting is described by a unitary two-mode squeezing operation \cite{VW06},
	\begin{equation}
		\label{eq:SqOp}
	\begin{split}
		&\hat S
		=\exp\left(\zeta^\ast\hat a\otimes\hat a-\zeta\hat a^\dag\otimes\hat a^\dag\right)
		\\
		=&\frac{1}{\mu}e^{-\nu\hat a^\dag\otimes\hat a^\dag/\mu}
		\left(\frac{1}{\mu}\right)^{\hat n}\otimes\left(\frac{1}{\mu}\right)^{\hat n}
		e^{\nu^\ast\hat a\otimes\hat a/\mu},
	\end{split}
	\end{equation}
	with the complex squeezing parameter $\zeta$, and the abbreviations $\mu=\cosh|\zeta|$ and $\nu=e^{i\arg\zeta}\sinh|\zeta|$, satisfying $\mu^2-|\nu|^2=1$.
	Note that $\zeta$ relates to the coherent amplitude of the optical pump---as well as coupling parameters and interaction time---for this process.
	The two involved modes are described through annihilation operators written as $\hat a\otimes\hat 1$ and $\hat 1\otimes\hat a$, similarly extending to the respective photon-number operators $\hat n\otimes\hat 1$ and $\hat 1\otimes\hat n$.

	The unitary $\hat S$ leads to the following transformations \cite{VW06}:
	\begin{subequations}
	\begin{eqnarray}
		\label{eq:TMSV}
		\hat S\left(|0\rangle\otimes|0\rangle\right) &=& |\lambda\rangle=\sqrt{1-|\lambda|^2}\sum_{n=0}^\infty \lambda^n |n\rangle\otimes|n\rangle,
		\\
		\label{eq:ModeOpAmp}
		\hat S\left(\hat a\otimes\hat 1\right)\hat S^\dag &=& \mu \hat a\otimes\hat 1+\nu\hat 1\otimes\hat a^\dag,
		\\
		\label{eq:ModeOpAmpVV}
		\hat S\left(\hat 1\otimes\hat a\right)\hat S^\dag &=& \mu \hat 1\otimes\hat a+\nu\hat a^\dag\otimes\hat 1,
	\end{eqnarray}
	\end{subequations}
	where $\lambda=-\nu/\mu=-e^{i\arg\zeta}\tanh |\zeta|$.
	The quantum state $|\lambda\rangle$ in Eq. \eqref{eq:TMSV} is typically referred to as a two-mode squeezed vacuum state.
	Furthermore, Eqs. \eqref{eq:ModeOpAmp} and \eqref{eq:ModeOpAmpVV} show that a signal in one mode is amplified by $\mu>1$ (for $|\zeta|>0$), and a seeding in the other mode is added coherently.
	Both effects are explored in more detail later (Secs. \ref{sec:Melanie} and \ref{sec:Lennart}).
	In addition, an intensity gain factor $\gamma$ can be defined as
	\begin{equation}
		\label{eq:GainFact}
		\gamma=\mu^2=\cosh^2|\zeta|=\frac{1}{1-|\lambda|^2}\geq 1.
	\end{equation}

\subsection{Seeded amplification and conditional measurements}

	Beyond this standard approach to squeezing operators, we can apply this nonlinear process to our exponential operators and partial traces.
	Using relations rigorously derived in Appendix \ref{app:OpAlg}, we can analytically describe how $\hat S$ acts on our exponential operators,
	\begin{equation}
	\begin{split}
		& \hat S\left[\hat E(x)\otimes\hat E(y)\right]\hat S^\dag
		\\
		=& \frac{1-|\lambda|^2}{1-|\lambda|^2xy}
		\exp\left(\frac{\lambda[1-xy]}{1-|\lambda|^2xy} \hat a^\dag\otimes\hat a^\dag\right)
		\hat E\left(
			\frac{x[1-|\lambda|^2]}{1-|\lambda|^2xy}
		\right)
		\\
		&  \otimes \hat E\left(
			\frac{y[1-|\lambda|^2]}{1-|\lambda|^2xy}
		\right)
		\exp\left(\frac{\lambda^\ast[1-xy]}{1-|\lambda|^2xy} \hat a\otimes\hat a\right).
	\end{split}
	\end{equation}
	Note that the above expression can be understood as a scenario in which two thermal states [cf. Eq. \eqref{eq:Thermal}] impinge on a two-mode squeezer, which corresponds to a second-order nonlinear interferometer.
	Furthermore, it is worth mentioning that coherence between the two modes is described via the terms $\exp\left(\lambda^\ast[1-xy] \hat a\otimes\hat a/[1-|\lambda|^2xy]\right)$ and its Hermitian conjugate.
	Again, we emphasize that derivatives with respect to $x$ and $y$ [cf. Eq. \eqref{eq:Photons}] enable us to describe photon states as inputs to the nonlinear interferometer.

	In addition to this general finding, we consider a measurement of $\hat E(z)$ in the second mode, e.g., for describing conditional measurements and heralding scenarios with an on-off detector.
	For this purpose, we perform a trace operation in the second mode while leaving the first mode untouched,
	\begin{equation}
	\begin{split}
		\label{eq:GeneralNL}
		& \mathrm{id}\otimes\mathrm{tr}\left[
			\left(\hat S\left[\hat E(x)\otimes\hat E(y)\right]\hat S^\dag\right) \left(\hat 1\otimes\hat E(z)\right)
		\right]
		\\
		=& \frac{1-|\lambda|^2}{1-y(|\lambda|^2x+[1-|\lambda|^2]z)}\hat E(\xi)
	\end{split}
	\end{equation}
	where ``$\mathrm{id}$'' denotes the identity and using the abbreviation
	\begin{equation}
		\label{eq:XiParameter}
		\xi {=} \frac{
			x(1{-}|\lambda|^2xy)(1{-}|\lambda|^2)
			-xzy(1{-}|\lambda|^2)^2
			+z(1{-}xy)^2|\lambda|^2
		}{
			(1-|\lambda|^2 xy)(1-|\lambda|^2xy-[1-|\lambda|^2]zy)
		}.
	\end{equation}
	See Appendix \ref{app:OpAlg} for technical details on the derivation of this exact formula of the partial trace.

\subsection{Special case}\label{subsec:SpecialCase}

\begin{figure}
	\includegraphics[width=.95\columnwidth]{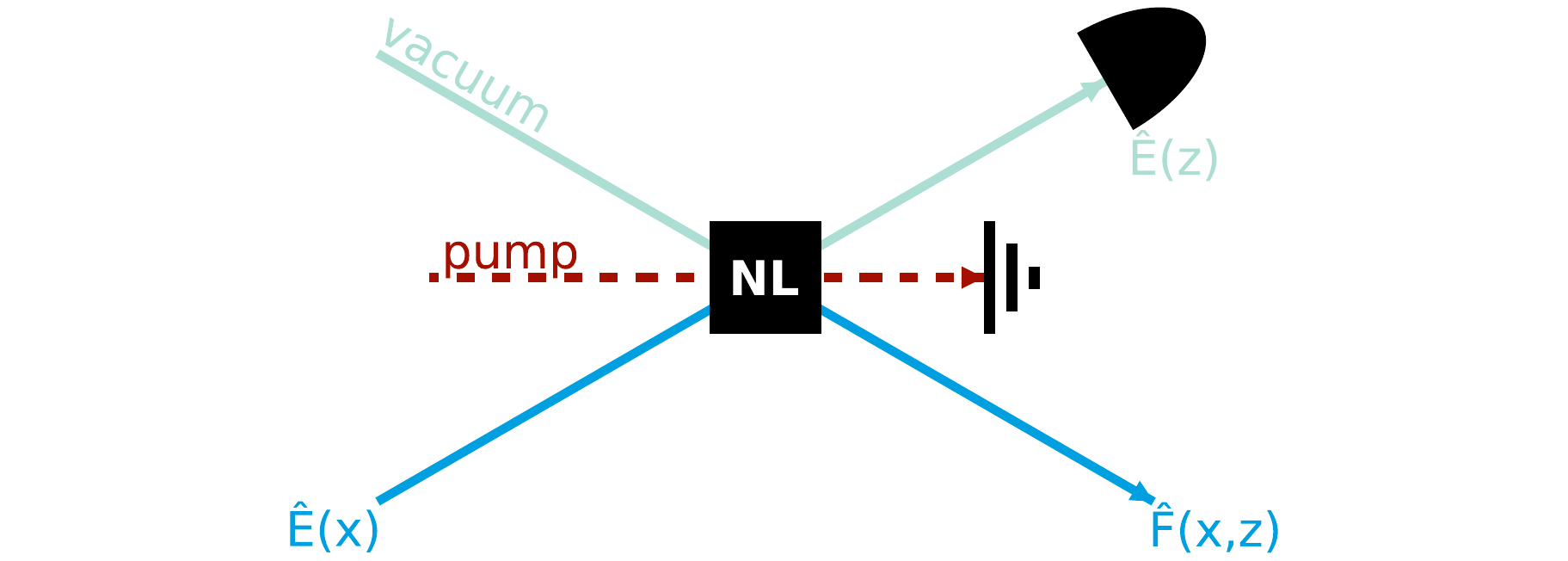}
	\caption{(Color online)
		Schematic depiction of the considered nonlinear interferometer, representing a two-mode squeezing operation (labeled ``NL'') driven by a pump.
		The lower input, resembling the mode $\hat a\otimes\hat 1$, is seeded with $\hat E(x)$, and vacuum is assumed for the upper input mode $\hat 1\otimes\hat a$.
		Measuring $\hat E(z)$ at the upper output leaves the operator $\hat F(x,z)$ from Eq. \eqref{eq:RelevantOutput} at the other output channel.
		Later, the input and output shall be merged to create a feedback loop.
	}\label{fig:SpecialCase}
\end{figure}

	In Fig. \ref{fig:SpecialCase}, we consider a special case of the previously derived expression in which we set $y=0$---meaning that $\hat E(y)=|0\rangle\langle 0|$ denotes the vacuum state.
	In this scenario, Eq. \eqref{eq:GeneralNL} simplifies to
	\begin{equation}
		\label{eq:RelevantOutput}
	\begin{split}
		\hat F(x,z)
		=& \mathrm{id}\otimes\mathrm{tr}\left[
			\left(\hat S\left[\hat E(x)\otimes|0\rangle\langle 0|\right]\hat S^\dag\right) \left(\hat 1\otimes\hat E(z)\right)
		\right]
		\\
		=&\frac{1}{\gamma}\hat E\left(\frac{x}{\gamma}+\frac{[\gamma{-}1]z}{\gamma}\right),
	\end{split}
	\end{equation}
	using the gain factor defined in Eq. \eqref{eq:GainFact}.
	The operator $\hat F(x,z)$, which can be fully expressed through $\hat E$, is important for our considerations in the continuation of this work.
	More specifically, we can represent the seeded nonlinear process via the input-output relation
	\begin{equation}
		\hat E(x)\mapsto \hat F(x,z),
	\end{equation}
	which depends on the conditioning $z$ and describes the interferometer shown in Fig. \ref{fig:SpecialCase}.

	For example, and based on Eq. \eqref{eq:RelevantOutput}, we can now directly compute the scenario in which $n$ photons are used to seed the nonlinear interferometer and a projection onto $m$ photons is performed in one output.
	Using Eq. \eqref{eq:Photons}, we directly find
	\begin{equation}
	\label{eq:Fnm}
	\begin{split}
		\hat F_{n,m}=& \mathrm{id}\otimes\mathrm{tr}\left(
			\hat S|n\rangle\langle n|\otimes |0\rangle\langle 0|\hat S^\dag
			\left[\hat 1\otimes|m\rangle\langle m|\right]
		\right)
		\\
		= &
		\frac{1}{n!m!}\left.\partial_{x}^n\partial_{z}^m\hat F(x,z)\right|_{x=0,z=0}
		\\
		= &
		\frac{(\gamma-1)^m}{\gamma^{\, m+n+1}}\frac{(m+n)!}{m!n!}|m+n\rangle\langle m+n|.
	\end{split}
	\end{equation}
	In relation to Fig. \ref{fig:SpecialCase}, this means that the lower input is seeded with an $n$-photon state and the conditional measurement at the upper output records $m$ photons.
	Then the lower output is given by $\hat F_{n,m}$ in Eq. \eqref{eq:Fnm}, showing that $m$ photons have been added to the initial $n$ photons with a success probability which corresponds to the scalar factor preceding $|n+m\rangle\langle n+m|$.

	Based on the methods presented so far, we have been able to rigorously model our nonlinear element with regards to arbitrary photon inputs and ideal projective measurements onto photon number states.
	However, this is still restricted to a single-pass scenario.
	Yet, we are going to demonstrate that this is already sufficient in order to describe feedback loops (i.e., multiple, subsequent passes through the active element) under realistic experimental conditions.

\section{Vector-type representation}\label{sec:Representation}

	After the exact derivation of the action of the nonlinear process on a photonic input state, we now divert to a practical decomposition.
	This enables us to develop an easily accessible toolbox to model all elements and processes which are relevant for our active feedback loops.

\subsection{State representation}

	Phase stability is a costly resource in experiments, thus it is reasonable to consider phase-averaged states, resulting in density operators $\hat\rho$ which are diagonal in the photon-number basis.
	Furthermore, photon-number states can be expressed via Eq. \eqref{eq:Photons} as derivatives of the operator $\hat E(x)$.
	Higher derivatives themselves can be described as the limit of linear combinations of a function via a difference quotient, $\partial_x^nf(x)=\lim_{\varepsilon\to0}[\sum_{j=0}^n\binom{n}{j}(-1)^{n-j}f(x+\varepsilon j)]/\varepsilon^n$.
	Thus, a density operator (being a compact operator) which is diagonal in the photon-number basis can be approximated with Eq. \eqref{eq:Photons} and an arbitrary precision $\varepsilon$ in terms of the following linear combination:
	$\sum_{n\in\mathbb N}p_n|n\rangle\langle n|\approx \sum_{j\in\mathbb N}\left[\sum_{n=j}^\infty p_n \binom{n}{j}(-1)^{n-j}/(\varepsilon^{n}n!)\right]\hat E(\varepsilon j)$.
	
	Consequently, the decomposition of density operators of the considered class of states reads
	\begin{equation}
		\label{eq:StateExpansion}
		\hat\rho=\sum_{k} P_k\hat E(x_k).
	\end{equation}
	It is then convenient to identify this density operator $\hat\rho$ with an array of pairs $[P,x]$ to represent each product $P\,\hat E(x)$ in the sum,
	\begin{equation}
		\label{eq:DOvec}
		\vec\rho=([P_k,x_k])_k.
	\end{equation}
	For example, the thermal state in Eq. \eqref{eq:Thermal} is represented through a single pair, $\vec\rho_\mathrm{th}=([1/(\bar n+1),\bar n{/}(\bar n+1)])$.
	It is noteworthy that, in all scenarios considered in this work, the representation of states $\hat\rho$ in terms of a finite vector $\vec \rho$ is exact and not an approximation as it would be in the most general scenario motivated above.

	Moreover, the above representation directly enables us to obtain the photon-number expansion of the state from Eq. \eqref{eq:StateExpansion} and the vector $\vec \rho$.
	That is, the $n$th photon-number probability---also establishing the fidelity $\mathcal F$ of $\hat\rho$ with an $n$-photon state---reads
	\begin{equation}
		\label{eq:Fidelity}
		\mathcal F(\hat\rho,|n\rangle\langle n|)=\mathrm{tr}(\hat\rho|n\rangle\langle n|)=\sum_{k}P_kx_k^{\,n},
	\end{equation}
	directly resulting from Eq. \eqref{eq:ExpExpansion}.
	Similarly to this overlap with photon-number states, we can express the $m$th normally ordered moments of photon-number operators as
	\begin{equation}
		\mathrm{tr}(\hat\rho {:}\hat n^m{:})=\sum_k P_k\frac{m! x_k^m}{(1-x_k)^{m+1}},
	\end{equation}
	using the properties of exponential operators, particularly, $\partial_w^m\hat E(w)|_{w=1}={:}\hat n^m{:}$ [cf. Eq. \eqref{eq:MainOperator}].
	This is, for example, useful to compute correlation functions exactly.

\subsection{Measurement representation}

	When proceeding as done for states, we obtain a similar representation of measurement operators that are diagonal in the photon number by writing
	\begin{equation}
		\label{eq:POVMvec}
		\hat\Pi=\sum_l \pi_l\hat E(w_l)
		\text{ via }
		\vec\Pi=([\pi_l,w_l])_l.
	\end{equation}
	For the purpose of computing expectation values, we can now evaluate the expectation value
	\begin{equation}
		\label{eq:InnerProd}
		\mathrm{tr}(\hat\rho\hat\Pi)=\sum_{k,l} \frac{P_k\pi_l}{1-x_kw_l}=(\vec\rho,\vec\Pi)
	\end{equation}
	by applying the properties of exponential operators.
	Therein, $(\vec\rho,\vec\Pi)$ defines an inner-product-type functional for $\vec\rho$ and $\vec\Pi$.

	A trivial example of a measurement operator is $\hat 1$, represented by $\vec\Pi=\vec 1=([1,1])$.
	The normalization of the state $\hat\rho$ as expanded in Eq. \eqref{eq:StateExpansion} is then obtained as $(\vec 1,\vec \rho)=\sum_{k}P_k/(1-x_k)$.
	If the state is properly normalized, this gives $(\vec 1,\vec \rho)=1$.
	If the state is obtained via conditional measurement (e.g., heralding) (cf. also Fig. \ref{fig:SpecialCase}), the quantity $(\vec 1,\vec \rho)$ resembles the success probability to produce this state.

	In Eq. \eqref{eq:OnOff}, the positive operator-valued measure of a single on-off detector is shown.
	More generally, one can consider a multiplexing detection scheme that consists of splitting a signal light field into $N$ modes with identical intensities and measuring each of those modes with an on-off detector separately \cite{PTKJ96}.
	In the ideal scenario, assuming unit efficiencies and vanishing dark-count rates, the measurement operator for obtaining $K\in\{0,\ldots,N\}$ clicks takes the form \cite{SVA12}
	\begin{equation}
		\label{eq:ClickCounting}
	\begin{split}
		\hat\Pi_K
		=& {:}
			\binom{N}{K}\left(e^{-\hat n/N}\right)^{N-K}\left(\hat 1-e^{-\hat n/N}\right)^{K}
		{:}
		\\
		=&\sum_{J=0}^K\binom{N}{K}\binom{K}{J}(-1)^{K-J}\hat E(J/N),
	\end{split}
	\end{equation}
	which is a (finite) linear combination of operators $\hat E(w)$ for $w\in\{0/N,1/N,\ldots,N/N\}$.
	This results in the vector representation $\vec \Pi_K=([\binom{N}{K}\binom{K}{J}(-1)^{K-J},J/N])_{J\in\{0,\ldots,N\}}$.
	This exact detector model already includes saturation effects, meaning the correct treatment of photon numbers that exceed the total number of detectors $N$.
	It is also worth mentioning that Eq. \eqref{eq:ClickCounting} converges to photon-number measurements for an infinite number of multiplexing steps and detectors, $\hat\Pi_K\to|K\rangle\langle K|$ for $N\to\infty$ \cite{SVA12}.

\subsection{General process representation}

	The third building block which is essential for the quantum description are processes.
	One can express each process in terms of the corresponding input-output relation,
	\begin{equation}
		\hat\rho\mapsto\Lambda(\hat\rho),
	\end{equation}
	where $\Lambda$ defines the quantum channel that models the evolution under study.
	From expectation values, $\mathrm{tr}(\Lambda[\hat\rho]\hat\Pi)=\mathrm{tr}(\hat\rho\Lambda^\dag[\hat\Pi])$, the known map for density operators implies the operation
	\begin{equation}
		\hat\Pi\mapsto \Lambda^\dag(\hat\Pi),
	\end{equation}
	which mathematically describes how the process acts on measurement operators.
	Therein, $\Lambda^\dag$ is the adjoint map to $\Lambda$, with respect to the Hilbert-Schmidt inner product.
	It is worth recalling that $\Lambda(\hat\rho)$ relates to the Schr\"odinger (i.e., state-based) picture of a process, and $\Lambda^\dag(\hat\Pi)$ defines the corresponding Heisenberg (i.e., measurement-based) picture.
	Furthermore, it is also noteworthy that the composition of a first process, $\Lambda'$, with a second one, $\Lambda''$, to get the overall process $\Lambda$ obeys
	\begin{equation}
		\Lambda(\hat\rho)=\Lambda''(\Lambda'(\hat\rho))
		\text{ and }
		\Lambda^\dag(\hat\Pi)={\Lambda'}^\dag({\Lambda''}^{\dag}(\hat\Pi)).
	\end{equation}
	This naturally extends to more than two operations and is convenient to consider powers of a single channel $\Lambda$ for representing multiple round trips in a loop configuration.

	Again, for our purposes, it is sufficient to describe the action on $\hat E$.
	For example, we may describe a loss channel with a quantum efficiency $\eta$.
	For a measurement, the loss is typically modeled as \cite{VW06}
	\begin{equation}
		\Lambda^\dag[\hat E(w)]={:}\exp[(w-1)\eta\hat n]{:}=\hat E(\eta w+1-\eta),
	\end{equation}
	implying that $\Lambda$ acts on measurement vectors in Eq. \eqref{eq:POVMvec} as
	\begin{equation}
		\label{eq:LossChannel}
		\Lambda^\dag(\vec\Pi)=([\pi_l,\eta w_l+1-\eta])_{l}.
	\end{equation}
	For determining the action of loss on a state, we can employ Eq. \eqref{eq:InnerProd}, yielding
	\begin{eqnarray*}
		& & \mathrm{tr}\left[\Lambda[\hat E(x)]\hat E(w)\right]
		=\mathrm{tr}\left[\hat E(x)\Lambda^\dag[\hat E(w)]\right]
		\\
		&=& \frac{1}{1-x[\eta w+1-\eta]}
		=\frac{1}{1-[1-\eta]x}\frac{1}{1-w\frac{\eta x}{1-[1-\eta]x}}.
	\end{eqnarray*}
	Thus, we find the adjoint operator $\Lambda$ to $\Lambda^\dag$, which reads
	\begin{equation}
		\label{eq:LossChannelState}
	\begin{split}
		\Lambda[\hat E(x)]
		=&\frac{1}{1-[1-\eta]x}\hat E\left(\frac{\eta x}{1-[1-\eta]x}\right),
		\\
		\Lambda(\vec \rho)
		=&\left(\left[\frac{P_k}{1-[1-\eta]x_k},\frac{\eta x_k}{1-[1-\eta]x_k}\right]\right)_k.
	\end{split}
	\end{equation}
	The latter expression shows the action of the loss channel $\Lambda$ on the state vector in Eq. \eqref{eq:DOvec}.
	In addition, it is straightforward to verify that the composition of two loss channels is described through a single loss channel with $\eta=\eta'\eta''$.

	Beyond losses, dark counts can be treated in a similar fashion, cf. Appendix \ref{app:Clicks}.
	However, for our types of detectors, the dark count rate is negligible \cite{BKSSV17}.
	Consequently, we set the dark count contribution to zero for the remainder of this work, $\delta=0$, and focus on the impact of more relevant imperfections.

\subsection{Nonlinear process representation}\label{sec:NLRep}

	A loss channel represents a passive element.
	The main focus in this work is, however, on active elements as analyzed in Sec. \ref{sec:ActiveElement}.
	Therein, we already derived that, for a conditioning to $\hat E(z)$, the input state $\hat E(x)$ maps to
	\begin{equation}
		\hat F(x,z)=\frac{1}{\gamma}\hat E\left(
			\frac{x+[\gamma-1]z}{\gamma}
		\right)
		=\Lambda[\hat E(x)].
	\end{equation}
	See also Fig. \ref{fig:SpecialCase}.
	Again, our inner product enables us to compute
	\begin{eqnarray*}
		&& \mathrm{tr}\left[\Lambda[\hat E(x)]\hat E(w)\right]
		=\mathrm{tr}\left[\hat E(x)\Lambda^\dag[\hat E(w)]\right]
		\\
		&=& \frac{1}{\gamma}\frac{1}{1-w\frac{x+[\gamma-1]z}{\gamma}}
		=\frac{1}{\gamma-[\gamma-1]zw}\frac{1}{1-x\frac{w}{\gamma-[\gamma-1]zw}},
	\end{eqnarray*}
	from which we obtain the impact on the measurement,
	\begin{equation}
		\Lambda^\dag[\hat E(w)]
		=\frac{1}{\gamma-[\gamma-1]zw}\hat E\left(
			\frac{w}{\gamma-[\gamma-1]zw}
		\right).
	\end{equation}
	Like for the case of loss, this can now be used to expand the action of the nonlinear channel onto the vectors for density operators $\vec\rho$ and measurement operators $\vec\Pi$.

	Since it is going to be of relevance, we also explicitly consider the special case $z=1$, resulting in an input-output formula for the nonlinear process in our notation,
	\begin{equation}
		\label{eq:AmpChannel}
	\begin{split}
		\Lambda(\vec\rho)
		=&\left(\left[\frac{P_k}{\gamma},\frac{x_k+[\gamma-1]}{\gamma}\right]\right)_k,
		\\
		\Lambda^\dag(\vec\Pi)
		=&\left(\left[\frac{\pi_l}{\gamma-[\gamma-1] w_l},\frac{w_l}{\gamma-[\gamma-1] w_l}\right]\right)_l.
	\end{split}
	\end{equation}
	This corresponds to the scenario in which one traces over the upper output in Fig. \ref{fig:SpecialCase}, $\hat E(0)=\hat 1$.
	Similar to the loss channel description, multiple processes of this amplifying form correspond to a single process, with, for example, $\gamma=\gamma'\gamma''$.

\subsection{Preliminary summary, limitations, and extensions}

	In summary, we formulated a vector-type formalism to easily access states [Eq. \eqref{eq:DOvec}] and measurements [Eq. \eqref{eq:POVMvec}], as well as the combination of both via a generalized inner product [Eq. \eqref{eq:InnerProd}].
	We also showed how this technique extends to processes, such as equipping the ideal click-counting operators [Eq. \eqref{eq:ClickCounting}] with losses [Eq. \eqref{eq:LossChannel}].
	Finally, we demonstrated that the nonlinear process depicted in Fig. \ref{fig:SpecialCase} can be straightforwardly embedded in this formalism [Eq. \eqref{eq:AmpChannel}].
	Moreover, we mentioned how success probabilities and the photon-number basis expansion follow from our vector representation.
	For practical purposes, it is particularly important to emphasize that the above findings enable us to implement a simple numerical toolbox for analyzing systems that include active elements.
	This is done by implementing the vector-based functions and relations and applying them---including arbitrarily complex combinations thereof---as needed.

	Our approach applies to any systems that are well described through the exponential-operator-based framework, using Eq. \eqref{eq:MainOperator}.
	This includes all states, operations and processes, and detection scenarios discussed previously.
	To further generalize our method, the central object of our studies can be modified.
	For example, a Kerr-type interaction---being quadratic in the photon-number operator---can be included by using the extended exponential operator ${:}\exp([x-1]\hat n+\tilde x\hat n^2){:}$, where the contribution proportional to the additional parameter $\tilde x$ accounts for the higher-order nonlinearity.
	Another example concerns two-mode scenarios, e.g., leading to operators ${:}\exp([x-1]\hat n\otimes 1+[y-1]\hat 1\otimes n+c\hat a\otimes\hat a^\dag+c^\ast\hat a^\dag\otimes\hat a){:}$, where the contributions for $c$ and $c^\ast$ relate to two-mode correlations from a beam splitter.
	Similarly, other nonlinear quantum effects and multimode scenarios can lead to significant future extensions of the fundamental framework introduced in this contribution.

	In the following section, we demonstrate the usefulness of the approach developed so far by applying it to two examples of experimental relevance, Secs. \ref{sec:Melanie} and \ref{sec:Lennart}.
	This includes not only the description of existing experiments but also the conception of future experiments with quantum light.

\section{Photon-number state generation}\label{sec:Melanie}

	Here, we apply our theoretical framework to model and improve state-of-the-art experiments to generate photon-number states.
	One aim of this description is to assess the expected quality of multiphoton states produced by repeated seeding of a PDC source of light and subsequent heralding.

\subsection{Motivation and setup description}

	Quantum metrology, quantum computation and communication, as well as fundamental studies of physics rely on the generation of complex quantum states \cite{D06,GT07,R17}, usually requiring single photons (e.g., for producing GHZ and W states \cite{DESBSP18}) and multiphoton states (e.g., Holland-Burnett states \cite{HB93,DZTDSW11} and cat states \cite{OJTG07}).
	Typical sources for these families of photonic states are single emitters (see, e.g., Refs. \cite{LD98,MKBSPZHI00,RSG19}) and PDC sources \cite{HABDMS13}.
	One possibility to enhance the performance of the latter kind of source are so-called quantum interference buffer \cite{MPDEQBBPS19}, relating to source multiplexing.
	Within this work, we focus on a dispersion-engineered PDC source as presented in Ref. \cite{HABDMS13}.

	Even if PDC sources are the workhorse in today's experiments, they have severe limitations in the generation probabilities of single- and multiphoton states, rendering this an outstanding problem \cite{TBHLNGS19}.
	This usually leads to the naive assumption that increasing the intensity of the pump pulse would solve the problem since the probability to generate $n$ photons increases with the mean photon number of the pump.
	But one encounters two main problems with this approach: an unreasonable power demand \cite{HBLNGS16} and unwanted higher photon-number components.
	The latter significantly diminishes the fidelity of the generated state with the target state \cite{TBHLNGS19}.
	Thus, a model of such contributions is essential to foresee the expected quality of produced states.

\begin{figure}
	\includegraphics[width=.95\columnwidth]{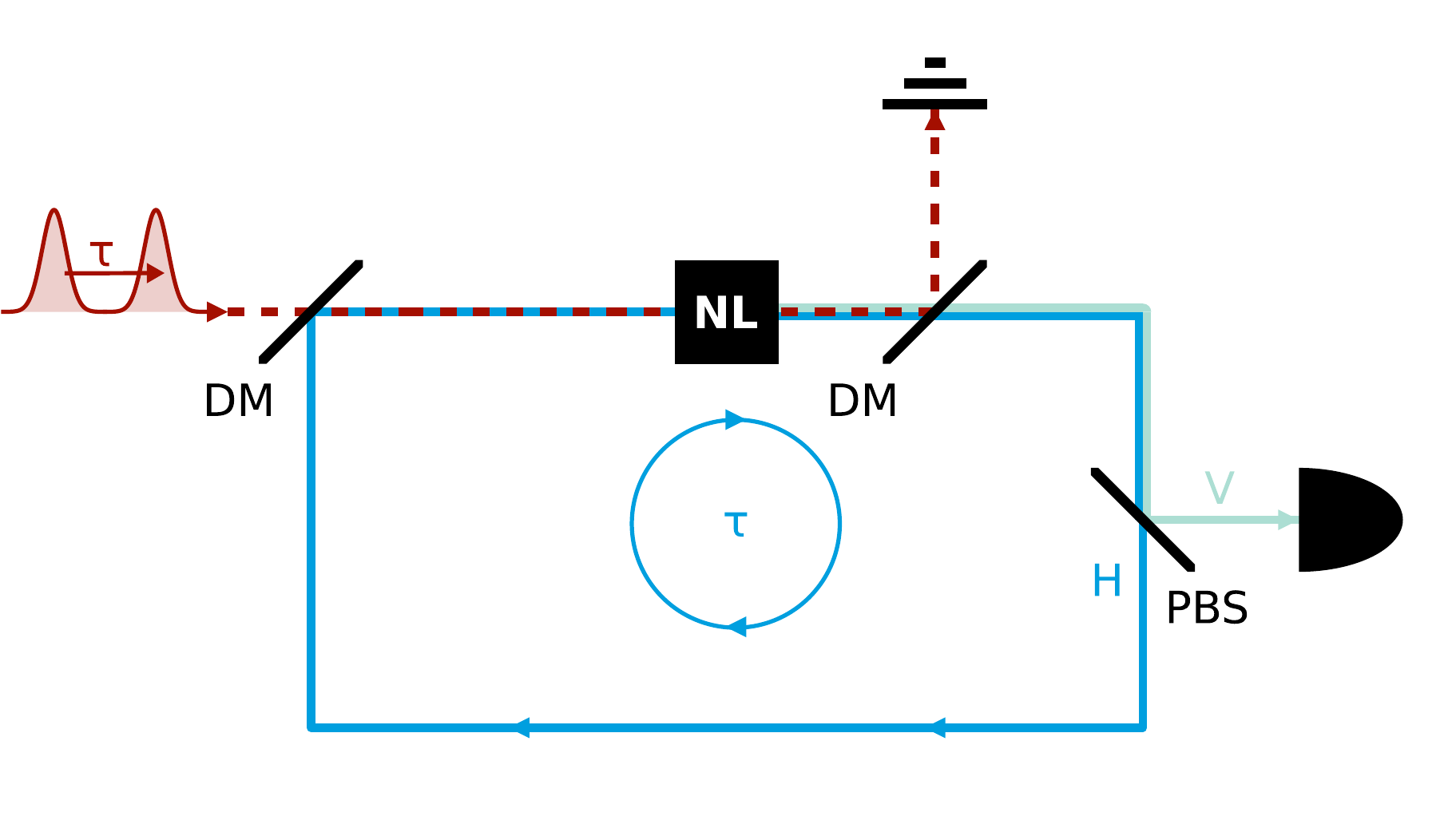}
	\caption{(Color online)
		Setup for higher-photon-number state generation.
		Laser pulses with a repetition rate $\tau$ optically pump a PDC source.
		The pump light is filtered at the second dichroic mirror (DM).
		The generated photon pairs in horizontal ($H$) and vertical ($V$) polarization propagate to a polarizing beam splitter (PBS).
		The vertically polarized light is reflected and measured with an on-off detector for heralding a photonic state, resembling a single photon.
		The horizontally polarized photons are transmitted and, therefore, enter the loop structure.
		This cycling mode overlaps with the subsequent pump pulse after passing the first DM again in the nonlinear element since the round trip time matches the repetition rate $\tau$.
		Because of this structure, the cycling mode stimulates the PDC process, acting as a seed to the process.
		If another click is reported at the detector, a second photon was heralded, i.e., coherently added to the cycling mode.
		This process can be repeated for multiple round trips in the loop.
	}\label{fig:Melanie}
\end{figure}

	In this section, we demonstrate how we mitigate the mentioned limitations by using active elements (i.e., time-multiplexed, pumped, and seeded PDC processes).
	One way to make use of active elements is to include a PDC source into time-multiplexing architecture, where we utilize quantum feedback; see Fig. \ref{fig:Melanie}.
	Time-multiplexing in this scheme makes use of generating photons in multiple time bins (defining temporal or pulse modes of light) to enhance the single- and multiphoton generation probabilities.
	Our PDC source generates polarization nondegenerate photon pairs.
	The horizontal polarization is sent into our time-multiplexing loop, which leads to a temporal overlap with the subsequent pump pulse, therefore serving as a feedback into the process.
	Thus, cycling photons induce self-stimulation, i.e., the seeded generation of the subsequent photon pairs.
	The vertical polarization is send to a detector.
	A click from this detector serves as an indicator of the successful stimulated generation of photons.
	The obtained photon-number state depends on the number $T$ of conditioning clicks and round trips in the loop.
	We have already shown the versatility of this setup because it can, in principle, produce complex quantum states of light since it enables us to generate tensor network states \cite{DESBSP18}.

\subsection{Modeling and characterization}

	For providing a theoretical model of the proposed iterative photon generation process in Fig. \ref{fig:Melanie}, we consider the scenario in Fig. \ref{fig:SpecialCase}, where $\hat E(z)$ is replaced by general $\hat\Pi=\sum_{l}\pi_l\hat E(z_l)$ and general input $\hat\rho=\sum_{k}P_k\hat E(x_k)$.
	Furthermore, we emphasize that the described optical mode is, in this case, the traveling mode in the loop configuration, showing that the applicability of our approach is not restricted to spatial modes but also extends to pulse modes.
	Using the approach in Sec. \ref{sec:NLRep}, the output state of this treatment then takes the form
	\begin{equation}
		\label{eq:MelInOut}
	\begin{split}
		\hat\rho \mapsto \hat\rho_{\hat\Pi}
		=&\sum_{k,l}P_k\pi_l\hat F(x_k,z_l),
		\text{ likewise}
		\\
		\vec\rho_\mathrm{out}=\Lambda(\vec\rho_\mathrm{in})
		=&\left(\left[
			\frac{P_k\pi_l}{\gamma},
			\frac{x_k+[\gamma-1]z_l}{\gamma}
		\right]\right)_{k,l}.
	\end{split}
	\end{equation}
	In the following, we first describe the setup in Fig. \ref{fig:Melanie} theoretically and then compare it to direct heralding techniques without feedback loops.
	In addition, our model enables us to study the influence of different imperfections separately, which is useful to distinguish different sources of experimental impurities.

	The above input-output relation \eqref{eq:MelInOut} describes a single seeded nonlinear process, stimulated with $\vec\rho_\mathrm{in}$, and a conditional measurement, expressed through a heralding with $\vec\Pi$.
	In the loop configuration in Fig. \ref{fig:Melanie} for $T$ round trips, $T$-fold application of $\Lambda$ has to be performed---meaning $\vec\rho_\mathrm{out}=\Lambda^T(\vec\rho_\mathrm{in})$.
	Note that, at this point, we have not included losses to focus on studying the the impact of the active element separately.
	Such imperfections are studied later by additionally including Eq. \eqref{eq:LossChannelState} in the loop and measurement description.

	To assess the quality of the produced states and rate of their production on a quantitative basis, two figures of merit are identified which are relevant in this context.
	First, the success probability $\mathcal P$ is given by the normalization of the resulting state,
	\begin{equation}
		\mathcal P
		=\frac{\mathrm{tr}(\hat\rho_{\hat\Pi})}{\mathrm{tr}(\hat\rho)}
		=\frac{(\vec \rho_\mathrm{out},\vec 1)}{(\vec \rho_\mathrm{in},\vec 1)},
	\end{equation}
	recalling that $\vec 1=([1,1])$ and even allowing for unnormalized inputs, $\mathrm{tr}(\hat\rho)=(\vec\rho_\mathrm{in},\vec 1)\neq 1$.
	Second, the fidelity in Eq. \eqref{eq:Fidelity}, normalized to $(\vec\rho_\mathrm{out},\vec 1)$, gives us the overlap of the output state with an $n$-photon state.

\begin{figure}
	\includegraphics[width=.95\columnwidth]{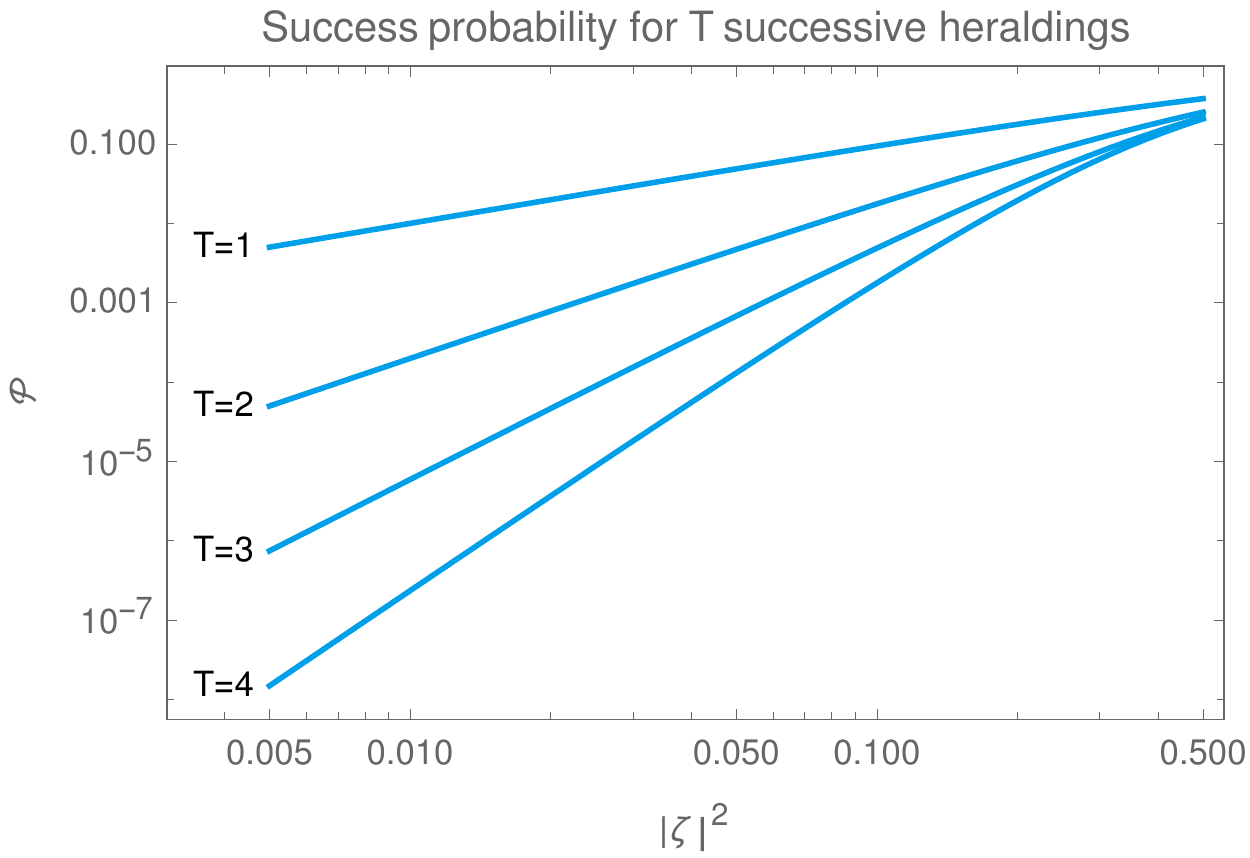}
	\\\vspace*{2ex}
	\includegraphics[width=.95\columnwidth]{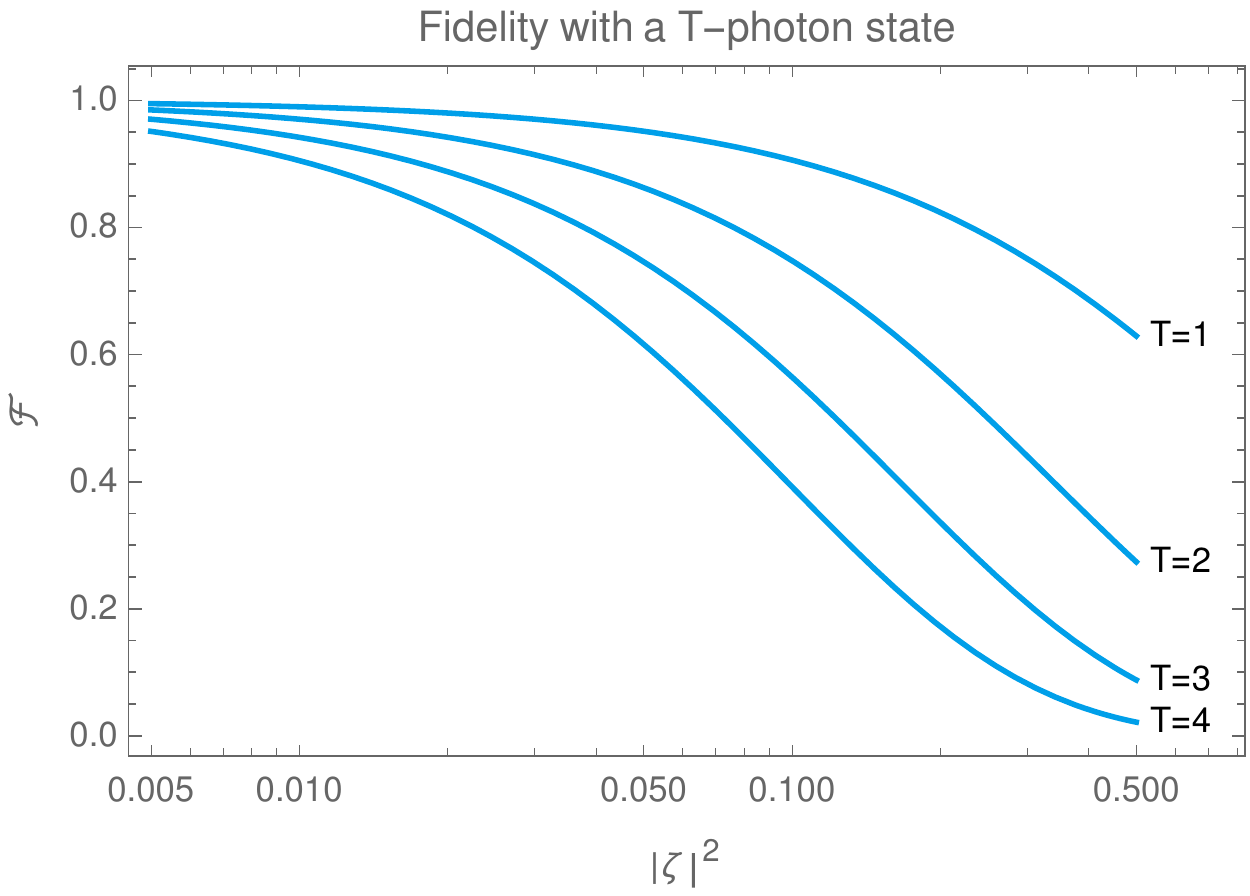}
	\caption{(Color online)
		Success probability (top) for $T$ successive heralding events and fidelity (bottom) with $|T\rangle\langle T|$ for the resulting state of the setup in Fig. \ref{fig:Melanie} for $T=1,2,3,4$ round trips when conditioned to a click of the on-off detector in each cycle.
		Both plots are shown as a function of the squared squeezing parameter $|\zeta|^2$, which is proportional to the pump power for the nonlinear process [Eq. \eqref{eq:SqOp}].
		The success probability---relating to the production rate of the state when scaled with $1/\tau$---increases with the pump power.
		At the same time, the fidelity with the target state decreases.
	}\label{fig:SuccProbFid}
\end{figure}

	Figure \ref{fig:SuccProbFid} shows the success probability $\mathcal P$ (top) and fidelity $\mathcal F$ (bottom) for $T$ round trips through the loop.
	The conditioning in each round trip is set to one click, $\hat\Pi=\hat 1-\hat E(1{-}\eta)$ [Eq. \eqref{eq:OnOff}], assuming a perfect detection efficiency $\eta=1$ and no losses when light propagates in the cycle.
	This idealized scenario is firstly investigated to assess the general possibility to produce multiphoton states $|T\rangle$ with the setup in Fig. \ref{fig:Melanie} with $T$ round trips.
	Both figures of merit ($\mathcal P$ and $\mathcal F$) are shown in Fig. \ref{fig:SuccProbFid} as a function of $|\zeta|^2=(\mathrm{arcosh}[\gamma^{1/2}])^2$ (being proportional to the pump intensity) on a logarithmic scale over two orders of magnitude.
	The success probability of heralding $T$ photons increases with $|\zeta|^2$ and is higher for lower $T$ values.
	The fidelity of the produced state with the targeted $T$-photon states increases with decreasing $|\zeta|^2$ values and is higher for smaller photon numbers $T$.

	The targeted photon-number states are nonclassical quantum states of a quantized radiation field \cite{MW95}.
	Thus, in order to quantify the nonclassical character of the actually produced states, we consider a method which is based on the moments of the click-counting statistics \cite{SVA13}, thus not requiring photon-number resolving detectors and additionally being exactly accessible within our framework and in experiments.
	For convenience, the constraints for classical light are briefly recapitulated in Appendix \ref{app:Clicks} and can be put into the form
	\begin{equation}
		\label{eq:SubBinomial}
		\mathcal N=\frac{\left(\mathrm{tr}[\hat\rho\hat E(1)]\right)\left(\mathrm{tr}[\hat\rho\hat E(0)]\right)}{\left(\mathrm{tr}[\hat\rho\hat E(1/2)]\right)^2}-1\geq 0,
	\end{equation}
	which can be expressed in terms of our inner products \eqref{eq:InnerProd}.
	A violation of this constraint certifies nonclassicality based on second-order correlation functions \cite{SVA13} and connects to the notion of sub-binomial light \cite{SVA12b,BDJDBW13}.

\begin{figure}
	\includegraphics[width=.95\columnwidth]{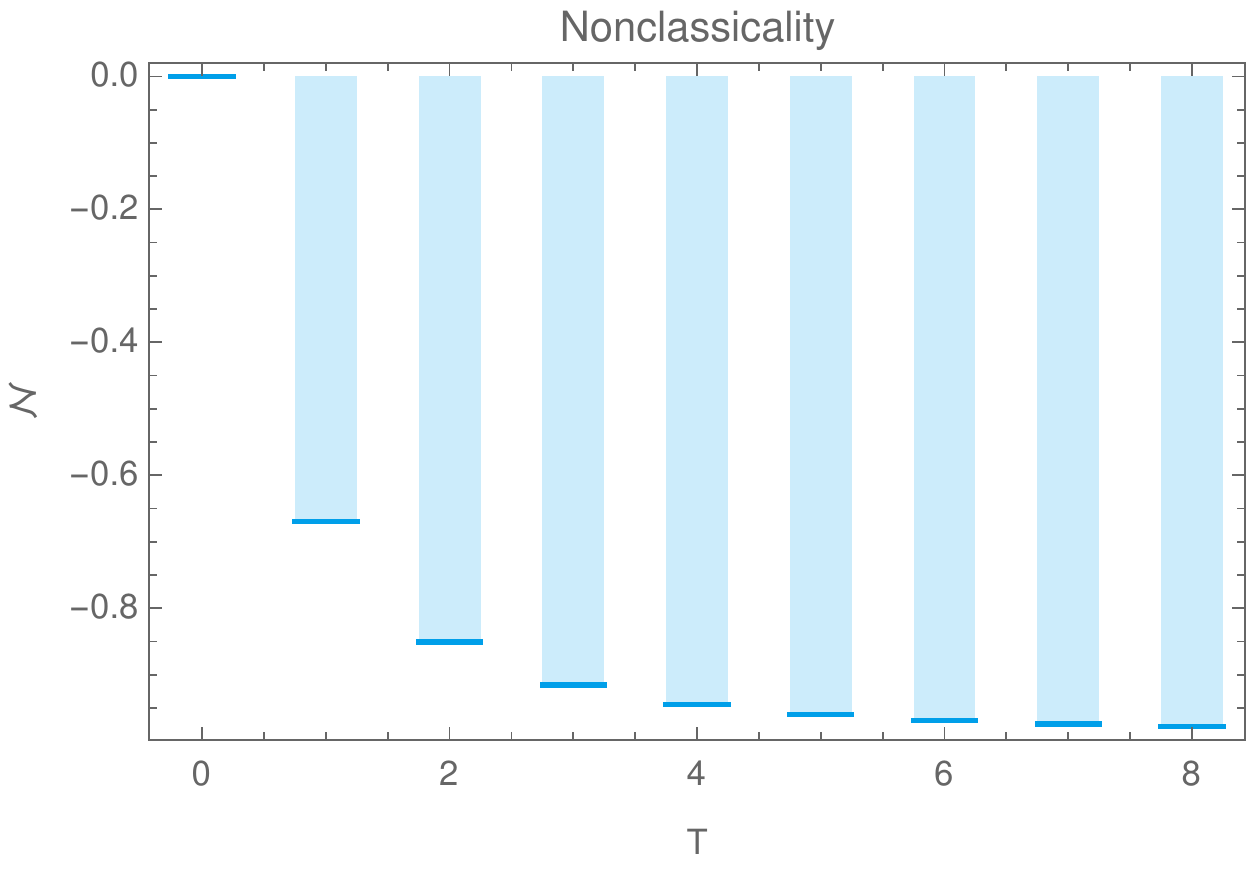}
	\caption{(Color online)
		Nonclassicality by violating inequality \eqref{eq:SubBinomial} as a function of the number $T$ of round trips and heralding events, resembling the active feedback-loop-based generation of $T$ photons.
		We set $|\zeta|^2=0.04$ (cf. Fig. \ref{fig:SuccProbFid}) and consider $1-\eta=10\%$ loss in the loop.
		The nonclassicality increases with $T$ and eventually saturates.
	}\label{fig:NclHeralding}
\end{figure}

	In Fig. \ref{fig:NclHeralding}, the nonclassicality, certified by violating inequality \eqref{eq:SubBinomial}, is shown for the state produced after the $T$th cycle.
	For those examples, we choose $\gamma=1.04$ (i.e., $|\zeta|^2\approx0.04$) and, in addition to the previous scenario, a round-trip loss of $10\%$.
	For $T=0$, we have a vacuum state, which must not show nonclassicality, $\mathcal N\geq0$.
	Hereafter, the nonclassicality increases with the number of heralded photons and saturates by converging to one.
	The latter behavior is in fact a result of the click-based nonclassicality condition which utilizes several on-off detectors, saturating for larger intensities.
	It might be important to emphasize that Fig. \ref{fig:NclHeralding} shows the nonclassicality of the iterative generation of up to eight photons and can be extended to any desired photon number by increasing the number of round trips and heralding events.

\begin{table}[b]
	\caption{
		Comparison of three heralding scenarios with respect to the success probability $\mathcal P$ of the heralding (second column) and the fidelity $\mathcal F$ with an ideal two-photon state (third column).
		We set the detection efficiency $\eta=80\%$ and $|\zeta|^2=0.04$.
		Scenario (i) describes two looped heralding processes, each conditioning to one click from a single on-off detector;
		scenario (ii) describes two looped heralding processes, each conditioning to one click from two multiplexed on-off detectors;
		and scenario (iii) describes a direct heralding process (no loop), conditioning to two clicks from two multiplexed on-off detectors.
		See Fig. \ref{fig:HeraldingCompare} for the resulting photon-number distributions.
	}\label{tab:Heralding}
	\begin{tabular}{p{.45\columnwidth}p{.25\columnwidth}p{.25\columnwidth}}
		\hline\hline
		Scenario & $\mathcal P$ & $\mathcal F$ \\
		\hline
		(i) & $1.94\text{\textperthousand}$ & $86.9\%$ \\
		(ii) & $1.85\text{\textperthousand}$ & $91.2\%$ \\
		(iii) & $0.49\text{\textperthousand}$ & $93.2\%$ \\
		\hline\hline
	\end{tabular}
\end{table}

\begin{figure}
	\includegraphics[width=.32\columnwidth]{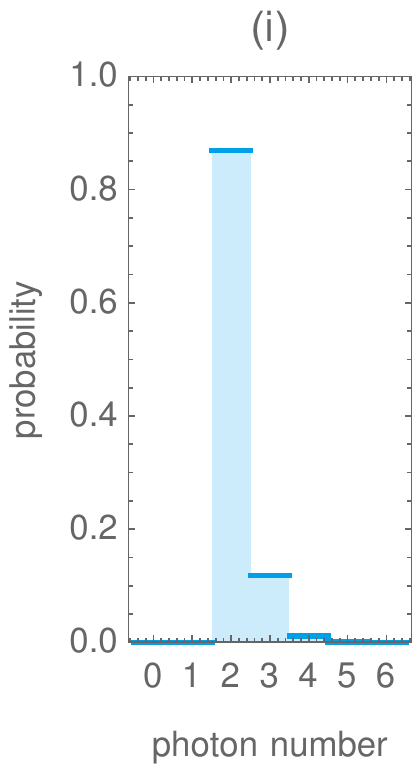}
	\includegraphics[width=.32\columnwidth]{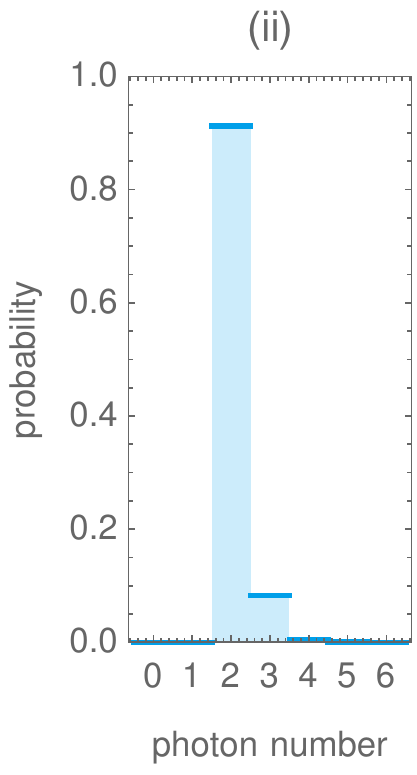}
	\includegraphics[width=.32\columnwidth]{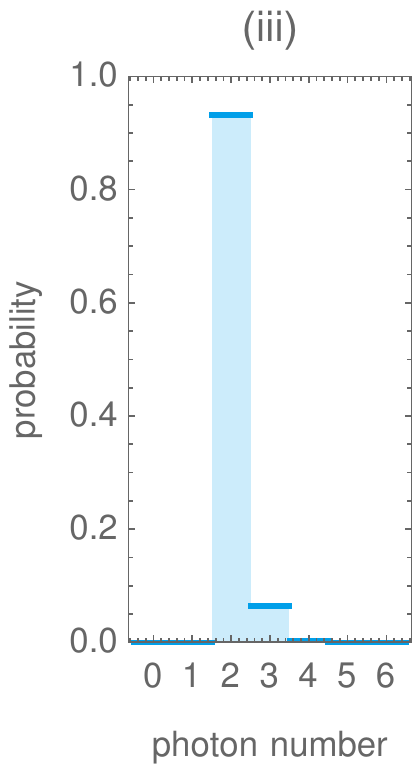}
	\caption{(Color online)
		Photon-number distribution of two-click heralded states for the different scenarios given in Table \ref{fig:HeraldingCompare}.
		From scenario (i) to (iii), left to right, the targeted probability of the two-photon component increases while higher (more than two) photon-number components are increasingly suppressed.
	}\label{fig:HeraldingCompare}
\end{figure}

	Finally, we may also compare our scheme with the direct heralding (i.e., without loop feedback) by employing click-counting devices consisting of $N$ multiplexed on-off detectors as our heralding measurement, Eq. \eqref{eq:ClickCounting}.
	The click-counting detector can result in $K=0,\ldots,N$ clicks.
	In addition, we assume a quantum efficiency of each on-off detector of $\eta=80\%$.
	We consider the following three scenarios (see Table \ref{tab:Heralding}):
	(i) light circles twice in loop, and a conditioning (i.e., heralding) to one click in each round trip from a single on-off detector is considered ($N=K=1$);
	(ii) light also travels twice in loop, but a conditioning to one click in each round trip from two multiplexed on-off detectors is considered ($N=2$ and $K=1$);
	and (iii) a direct heralding is considered by a conditioning to two clicks from two multiplexed on-off detectors ($N=K=2$), without a round trip.
	Scenario (iii) represents the commonly applied approach to produce a two photon state, scenario (i) uses our loop architecture as it is (Fig. \ref{fig:Melanie}), and scenario (ii) presents a combination of both previous approaches.
	In addition, in Fig. \ref{fig:HeraldingCompare}, the exact photon-number distribution for all three possibilities is depicted, for $\gamma=1.04$ and a nonunit detection efficiency, $\eta=80\%$, while now ignoring round-trip losses since they have been considered previously.

	Our analysis of all cases (Table \ref{tab:Heralding} and Fig. \ref{fig:HeraldingCompare}) shows that, for scenario (i), the success probability is comparably high, but the fidelity is comparably low.
	Conversely, the success probability is comparably low, yet the fidelity is comparably high in scenario (iii).
	Interestingly, scenario (ii) offers both a comparably high success probability and high fidelity.
	This demonstrates that a combination of direct and loop-based heralding schemes is in fact advantageous for experimentally producing higher-order photon-number states, going beyond existing schemes which employ multiplexing layouts to experimentally produce higher photon-number states \cite{SECMRKNLGWAV17,TMNBBS19}.
	As mentioned before, our method is not restricted to the specific number of photons considered here and can be scaled up easily to any photon number by increasing the number of round trips and the number of multiplexing detectors.
	Therefore, our theoretical model enables us to improve our initial setting in Fig. \ref{fig:Melanie} by replacing the single on-off heralding detector with a multiplexing detector to enhance the setup's performance in future multiphoton-generation experiments.
	Also note that the direct higher-order photon-number state heralding [scenario (iii)] is, by construction, limited to the number $N$ of available on-off detectors, which is not the case for our feedback-loop-based approach [scenarios (i) and (ii)].

\section{Balancing loss through amplification}\label{sec:Lennart}

	In this section, we apply our methodology to a second example of practical relevance.
	The purpose of this study is to compensate for losses, originating from the propagation in a loop, by means of amplification as commonly done in classical optics.
	However, quantum models of such amplifiers also introduces additional noise (see, e.g., Ref \cite{SVA14}), which has to be characterized for an optimal utilization of the amplifier.

\subsection{Motivation and setup description}

	Quantum walks in Mach-Zehnder setups are proven to provide a versatile platform to approach the goal of realizing a universal quantum simulator \cite{SCPGMAJS10,SCPGJS11,SGRLSPHJS12,NENGJBS16,NBKSSGPKJS18}.
	We devised a looped Michelson interferometer as a platform for time-multiplexed quantum walks, which overcomes some of the restrictions of previous implementations \cite{LMNPGBJS19}.
	The main advantage of this architecture is a higher-dimensional internal state for walkers (i.e., the photons), arising from the additionally available traveling direction in the loop, clockwise and counterclockwise.
	In addition, we successfully implemented three electro-optic modulators (EOMs) in our setup to manipulate the polarization of the traveling photons.
	This increased configurability allowed us to study the walker's evolution on complex graph structures, such as realizing quantum walks on a circle with periodic boundary conditions and other scenarios which are only accessible with higher-dimensional internal states \cite{LMNPGBJS19}.

\begin{figure}
	\includegraphics[width=.95\columnwidth]{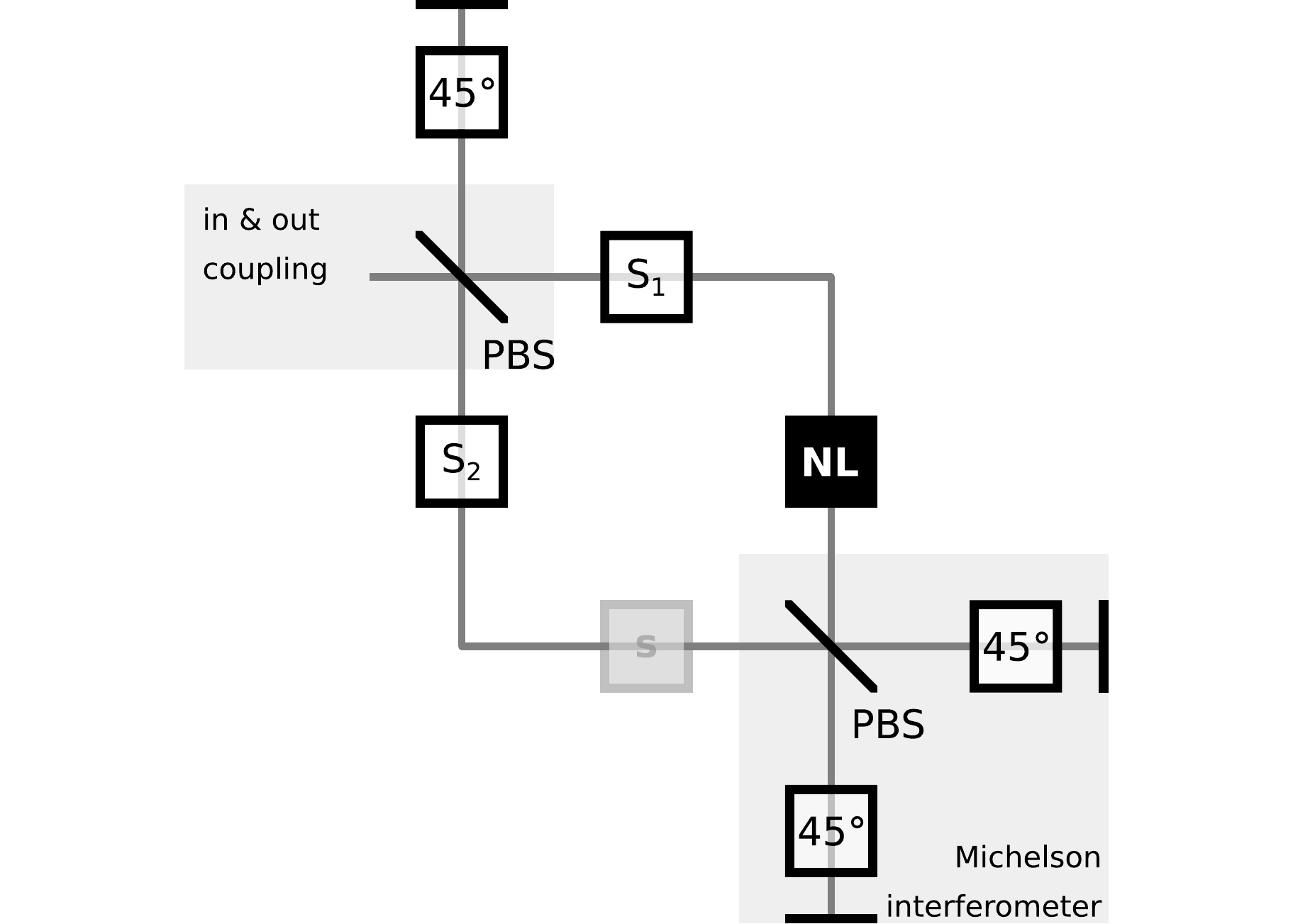}
	\caption{(Color online)
		Outline of a Michelson interferometer in a loop configuration with deterministic in- and out-coupling; see also Fig. \ref{fig:Lennart2}.
		An erbium-doped fiber amplifier serves as our active element within the main loop.
		In each interferometer arm, as well as in the second arm of the coupling stage, a $45^\circ$ polarization rotation together with a mirror (thick horizontal and vertical lines) reflect the incident light while simultaneously swapping horizontal with vertical polarization.
		Both switches $S_1$ and $S_2$, implemented as EOMs, enable us to further manipulate the polarization for the clockwise and counterclockwise traveling pulses of light; see also Fig. \ref{fig:Lennart2} in this context.
		A third switch $S$ (light gray) could modulate the mixing ratio of light at the PBS of the Michelson interferometer but is not used here.
	}\label{fig:Lennart1}
\end{figure}

\begin{figure}
	\includegraphics[width=.95\columnwidth]{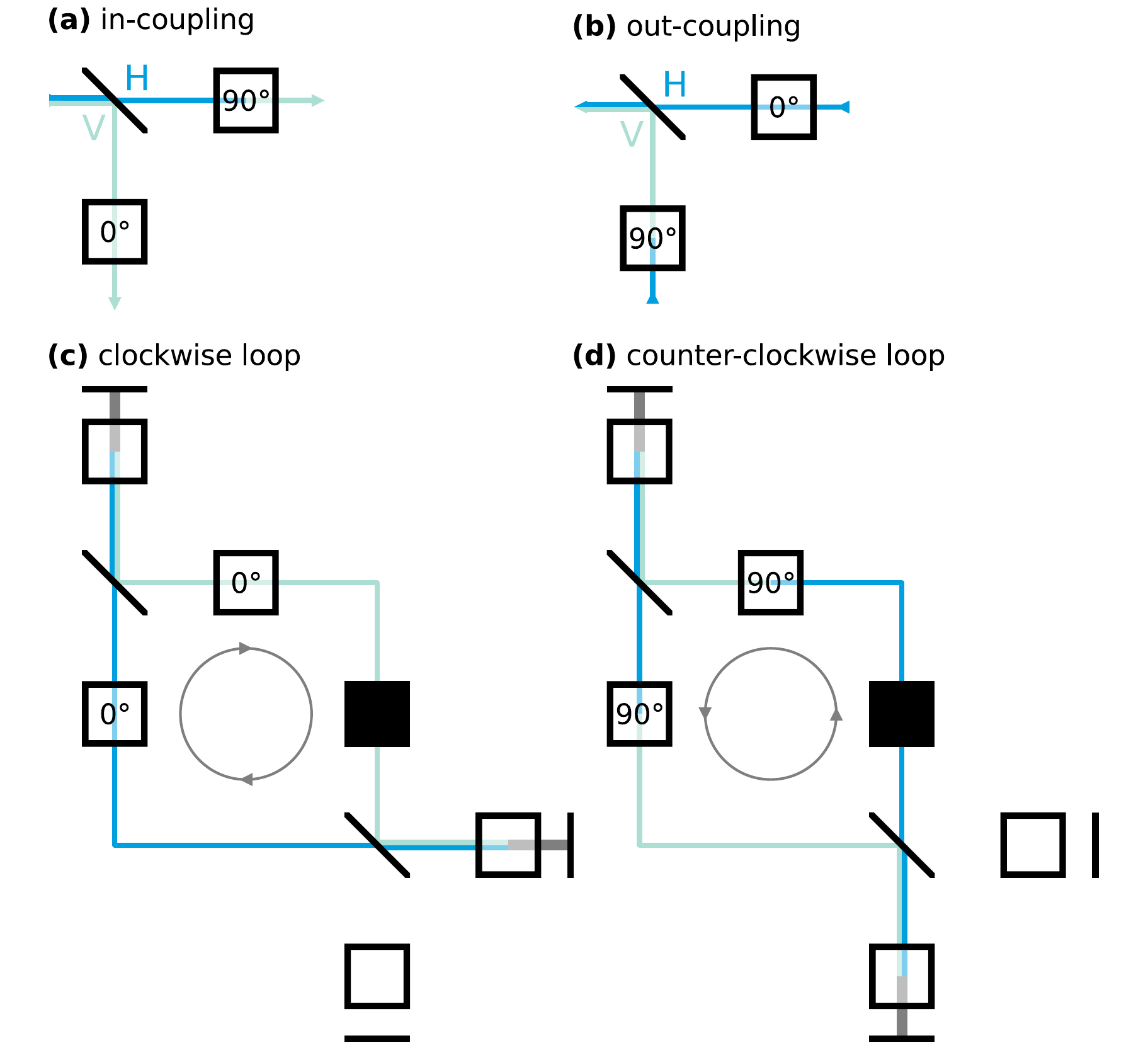}
	\caption{(Color online)
		Settings of switches $S_1$ and $S_2$ for the setup in Fig. \ref{fig:Lennart1}.
		For in- and out-coupling of type-II PDC light, being a two-mode squeezed vacuum state in polarization modes, one of the two switches is set to $0^\circ$, and the other one is set to $90^\circ$, cf. top panels (a) and (b).
		The paths of light pulses propagating in clockwise and counterclockwise direction are depicted in the bottom row, panels (c) and (d), respectively, together with the angles of polarization rotation implemented by the switches.
		Recall that, in the arm of the interferometer and coupling stage (cf. Fig. \ref{fig:Lennart1}), light passes twice through a $45^\circ$ rotation stage (before and after the reflection at the mirror), resulting in a swap of polarization.
	}\label{fig:Lennart2}
\end{figure}

	Beyond previously existing experiments, a revised version of this setup is described in Figs. \ref{fig:Lennart1} and \ref{fig:Lennart2}.
	The modifications mainly concern the introduction of an active element, together with an additional in- and out-coupling stage.
	The nonlinear component in this scenario is an erbium-doped fiber amplifier.
	The idea behind introducing this nonlinear optical element is to counter the losses in our setup, significantly impacting the quantum properties of light propagating in a feedback loop.
	See Refs. \cite{PCWZMTGJ18,ZJ20} for promising applications of this approach.
	Again, a comprehensive model of the quantum properties of the amplifier is required to assess and quantify the potential success of the proposed setup.

\subsection{Modeling and characterization}

	For studying amplification, we again apply the results in Eq. \eqref{eq:RelevantOutput} from Sec. \ref{subsec:SpecialCase}.
	While we pursued a state-based approach in the previous application, let us focus on a measurement-operator-based approach to complement our earlier considerations.
	The goal is to analyze the evolution of the quantum correlations from a two-mode squeezed vacuum state when including amplifiers to counter propagation losses in the Michelson interferometer loop.
	Before doing so, we analyze the excess noise from the nonlinear process to determine different possibilities for how one could overcome losses with amplification in the quantum domain.

	For simplicity, we begin by considering a thermal state with mean photon number $\bar n_\mathrm{in}$ [cf. Eq. \eqref{eq:Thermal}].
	Note, however, that the fundamental noise effects of amplification are not dependent on the specific input state.
	When including losses in the loop, Eq. \eqref{eq:LossChannelState} for $0\leq \eta\leq 1$, we get the reduced output photon number $\bar n_\mathrm{out}=\eta\bar n_\mathrm{in}$.
	We can also apply the amplification in Eq. \eqref{eq:AmpChannel}, which yields $\bar n_\mathrm{out}=\gamma\bar n_\mathrm{in}+(\gamma-1)$.
	The gain factor $\gamma\geq 1$ in Eq. \eqref{eq:GainFact} describes the amplification, and the latter summed term $\gamma-1$ is the excess noise of this process;
	see also Ref. \cite{SVA14} for further details.

	Combining first loss with second amplification, and iterate those processes $T$ times in a loop, we get
	\begin{equation}
		\bar n_\mathrm{out} =
		\left\lbrace
		\begin{array}{ll}
			\bar n_\mathrm{in}+T(\gamma-1)
			& \text{for }\eta\gamma=1,
			\\
			(\gamma\eta)^T\bar n_\mathrm{in}+\frac{1-(\gamma\eta)^T}{1-(\gamma\eta)}(\gamma-1)
			& \text{for }\eta\gamma\neq1.
		\end{array}\right.
	\end{equation}
	To exactly compensate for the losses, we can choose $\gamma\eta=1$.
	In this case, however, we also expect an additional noise contribution of $\gamma-1$ for each round trip.
	If we set $\gamma=(1+\bar n_\mathrm{in})/(1+\eta\bar n_\mathrm{in})$, resulting in $\gamma\eta=1-(1-\eta)/(1+\eta\bar n_\mathrm{in})<1$ for $\eta\neq1$, we get $\bar n_\mathrm{out}=\bar n_\mathrm{in}$.
	This constitutes what we define as the balanced scenario.
	Recall that excess noise is relatively small in the classical high-intensity regime when compared with the signal, rendering the balanced scenario an option that is mostly relevant in the quantum domain.
	Both cases (compensated and balanced) can be compared with the nonamplified propagation in the loop, i.e., $\gamma=1$.
	
	For characterizing the setup in Fig. \ref{fig:Lennart1}, we assume that a continuous-variable two-mode squeezed vacuum $|\tilde\lambda\rangle$ [Eq. \eqref{eq:TMSV}] enters the loop, where the two modes correspond to two polarizations.
	Again, other states of nonclassical light could be used similarly, such as the discrete-variable heralded photon-number states considered in the previous section.
	Considering broadly accessible sources, it makes sense, however, to focus on the specific example under study.
	The initially horizontal and vertical ($H$ and $V$) photons of the two-mode squeezed vacuum state propagate in a clockwise and counterclockwise direction through the setup, respectively; see Fig. \ref{fig:Lennart2}.
	After $T$ cycles, including loop losses and amplification acting separately on each polarization, the light pulses are coupled out and measured.
	Following a Heisenberg-picture-like approach, the measurement operators for horizontal and vertical light are propagated backwards according to Eqs. \eqref{eq:LossChannel} and \eqref{eq:AmpChannel}.
	The relevant integral to describe expectation values is then given by
	\begin{equation}
		\label{eq:TMSVExpect}
		\langle\tilde\lambda|\hat E(w)\otimes\hat E(w')|\tilde\lambda\rangle=\frac{1-|\tilde\lambda|^2}{1-|\tilde\lambda|^2ww'}
	\end{equation}
	for the two-mode squeezed input state $|\tilde\lambda\rangle$.
	More generally, a functional on the propagated measurement operators, represented via $\vec\Pi$ and $\vec\Pi'$, can be defined through the bilinear-form-like expression
	\begin{equation}
		\label{eq:BiLinForm}
		|\vec\Pi,\vec\Pi'|_{\tilde\lambda}=\sum_{l,l'}\frac{(1-|\tilde\lambda|^2)\pi_l\pi'_{l'}}{1-|\tilde\lambda|^2w_lw'_{l'}}.
	\end{equation}
	Note that the first derivative of the expression \eqref{eq:TMSVExpect} for $w$ and $w'$ at the value one yields the mean photon number for the horizontal and vertical component of the state $|\tilde\lambda\rangle$, $\mathrm{tr}\otimes\mathrm{tr}(\hat\rho[\hat 1\otimes\hat n])=\mathrm{tr}\otimes\mathrm{tr}(\hat\rho[\hat n\otimes\hat 1])=|\tilde\lambda|^2/(1-|\tilde\lambda|^2)$.
	Finally, to characterize nonclassical correlations between the $H$ and $V$ polarization, we can consider a click-based cross-correlation criterion to assess the quantum correlations between the polarizations \cite{SVA13,SBVHBAS15}; see also Appendix \ref{app:Clicks} for some details.

\begin{table}
	\caption{
		Amplification in the loop in Fig. \ref{fig:Lennart1} for four scenarios (first column).
		The first row shows the properties of the initial state $|\tilde\lambda\rangle$, with $|\tilde\lambda|^2=0.1$.
		The following rows include one round trip in the loop, including a loss of $1-\eta=20\%$.
		In those cases, the amplification is set to $\gamma=1$, $\gamma=1/\eta$, and $\gamma=[1-(1-\eta)|\tilde\lambda|^2]^{-1}$ to represent the nonamplified, loss compensated, and balanced [i.e., input equals output mean photon number (see singles)] scenarios.
		``Singles'' (column two) denotes the probability to measure a click for $H$ polarization (identical value for $V$), assuming a single on-off detector with unit efficiency.
		``Coincidences'' (column three) defines the probability for a click from each polarization.
		Nonclassicality, i.e., quantum correlations between $H$ and $V$, is certified through a negative value in the last column.
		See Fig. \ref{fig:TMSVcycles} for multiple round trips.
	}\label{tab:ClickCorr}
	\begin{tabular}{p{.25\columnwidth}p{.15\columnwidth}p{.225\columnwidth}p{.3\columnwidth}}
		\hline\hline
		Scenario & Singles & Coincidences & Cross-correlations
		\\\hline
		initial & $0.100$ & $0.100$ & $-0.804\times10^{-3}$
		\\
		no amplification & $0.082$ & $0.067$ & $-0.337\times10^{-3}$
		\\
		compensated & $0.265$ & $0.109$ & $+0.061\times10^{-3}$
		\\
		balanced & $0.100$ & $0.068$ & $-0.346\times10^{-3}$
		\\
		\hline\hline
	\end{tabular}
\end{table}

	In Table \ref{tab:ClickCorr}, we consider the initial state and the output states after one round trip in the loop shown in Figs. \ref{fig:Lennart1} and \ref{fig:Lennart2} for different amplification scenarios.
	For no amplification, $\gamma=1$, the losses of the propagating pulse diminishes the single counts and coincidence counts between the polarizations when compared with the initial state.
	When setting $\gamma\eta=1$, we completely compensate losses by a corresponding amplification factor.
	While the coincidences increase, the singles increase even more because of the unavoidable excess noise of the amplifier.
	In the balanced scenario, in which the singles are kept constant, the coincidences slightly increase.
	Note that on-off detectors are assumed to have a unit quantum efficiency to only study the impact of the loss in the loops and strategies to counter those imperfections through quantum amplifiers.

	Most significant is the impact of the different amplifications on the nonclassical cross-correlations, which have to be negative to certify nonclassicality \cite{SVA13,SBVHBAS15} (cf. also Appendix \ref{app:Clicks}).
	Here, the initial nonclassicality is significantly reduced by the loop losses; see the last column of Table \ref{tab:ClickCorr} for the nonamplifying scenario.
	Again, the strong contribution of excess noise affects the correlations, which are no longer detectable for $\gamma\eta=1$ because of the positive cross-correlation value.
	Conversely, the balanced amplification does slightly increase the verified nonclassical feature when compared with the nonamplified scenario, thus showing a much better performance than the compensated amplification case.

\begin{figure}
	\includegraphics[width=.95\columnwidth]{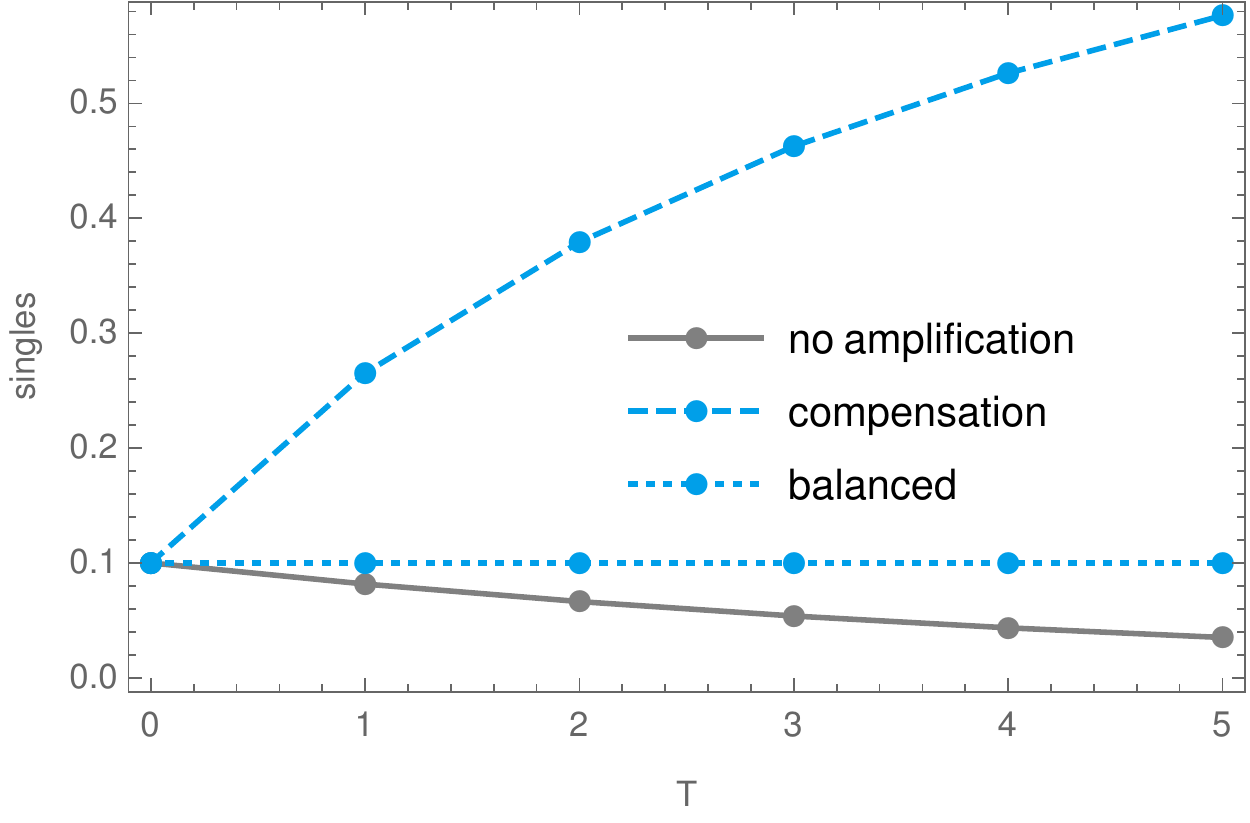}
	\includegraphics[width=.95\columnwidth]{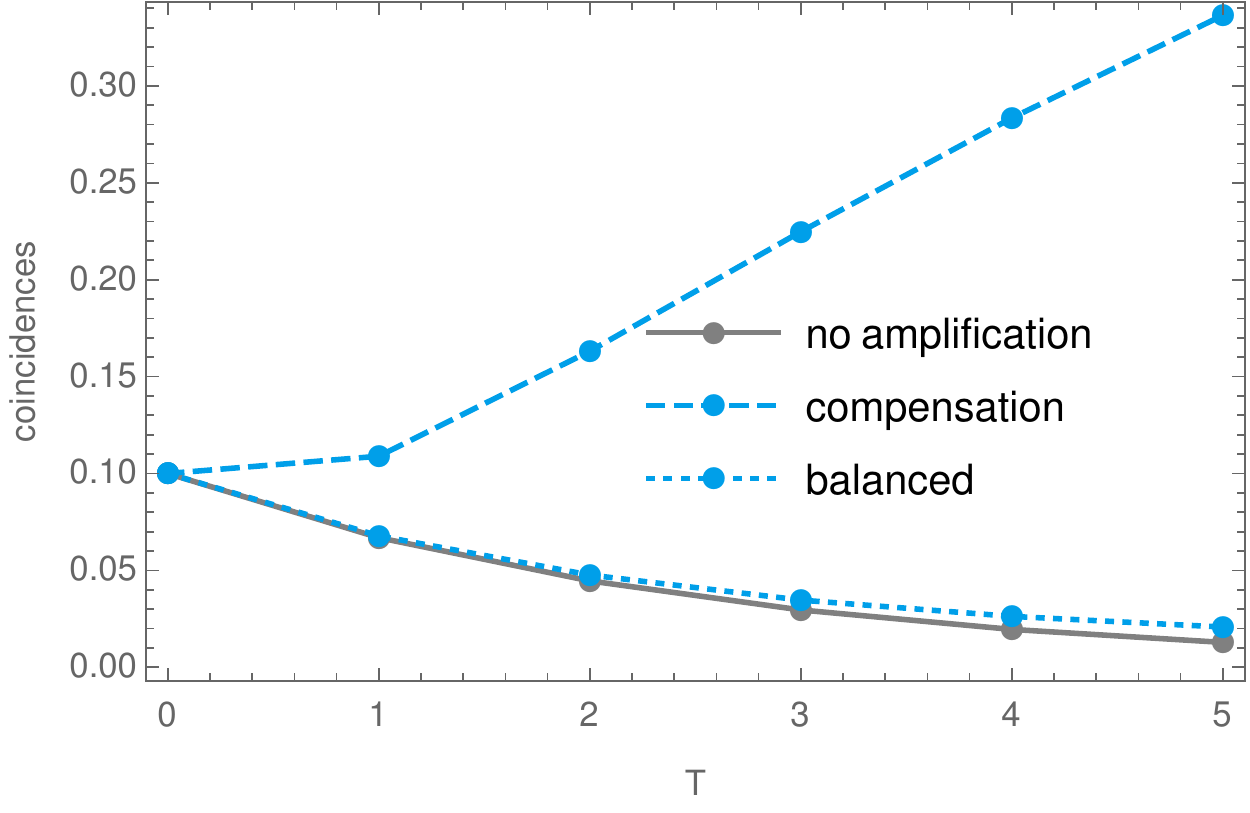}
	\includegraphics[width=.95\columnwidth]{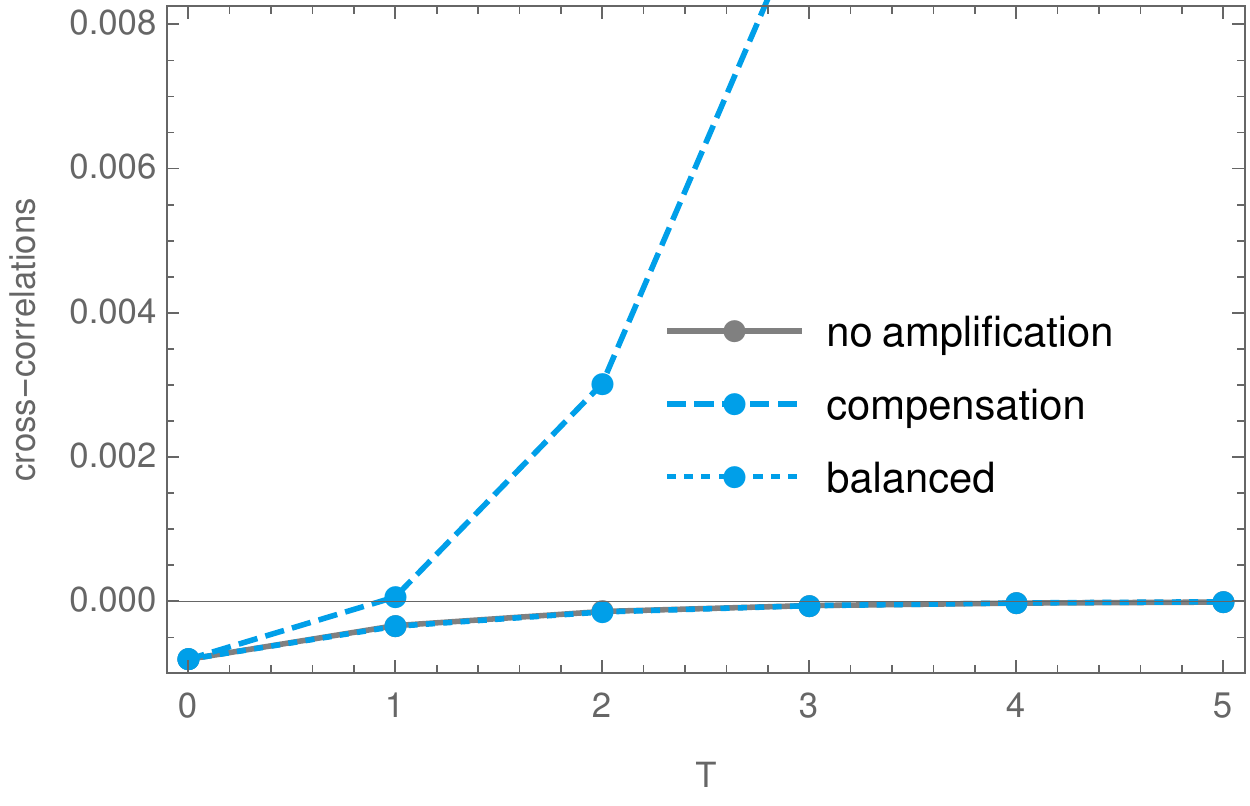}
	\caption{(Color online)
		Properties of the two-mode squeezed vacuum state $|\tilde\lambda\rangle$, with $|\tilde\lambda|^2=0.1$, propagating $T$ cycles in the loop setup in Fig. \ref{fig:Lennart1}, introducing $20\%$ loss per round trip.
		Three scenarios of amplifications are considered, $\gamma=1$ (no amplification, solid), $\gamma=1/\eta$ (compensation, dashed), and $\gamma=[1-(1-\eta)|\tilde\lambda|^2]^{-1}$ (balanced case, dotted).
		The top, middle, and bottom plots show the single counts for $H$ and $V$ (identical values), coincidence count probability, and the cross-correlations to infer nonclassicality, respectively.
	}\label{fig:TMSVcycles}
\end{figure}

	The evolution of the described features over multiple round trips $T$ is further analyzed in Fig. \ref{fig:TMSVcycles}.
	The top plot shows how single counts for the three considered amplification values vary.
	It can be seen that the balanced case undoes the effect of loss for all $T$ values.
	The middle plot also shows that the coincidences for the compensated case slightly exceed those for the pure loss case.
	In the nonclassical cross-correlations, both the nonamplifying and the balanced scenario exhibit nonclassicality, which is decreasing with $T$.
	By contrast, the compensated case fails to exhibit nonclassicality already after a single cycle in the loop, as indicated by the positive cross-correlation value.
	This is surprising since the commonly expected case to counter losses through amplification would be $\eta\gamma=1$, the compensation case.
	Yet, excess noise---correctly included in our methodology---spoils this expectation and, in fact, favors the balanced scenario.
	Again, our rigorous model enables us to find unexpected experimental situations (here, the balanced scenario) which are advantageous when compared with commonly applied usage of a nonlinear element (i.e., the standard loss compensation).

\section{Discussion and conclusion}\label{sec:Conclusion}

	In summary, we derived an exact theoretical framework for the unified description of loop-based optical networks which include a second-order nonlinear medium.
	By using a vector-type representation, we additionally formulated a numerical toolbox for implementing and applying our analytical findings.
	We then modeled two realistic experimental setups for realizing iterative photon-addition protocols and quantum amplification processes in photonic systems with looped optical paths.
	Our techniques enabled us to reveal nonintuitive improvements for both experiments, which highlight the power of our approach.

	Our general method is based on linear combinations of exponential operators of the photon-number operator, which is---as we demonstrated---already sufficient to cover a broad range of experiments.
	The resulting vector-type description describes density and measurement operators, and mappings between the exponential operators can be used to model all processes that occur in the scenarios we studied.
	Even the seeded nonlinear process with a conditional measurement can be modeled in this manner.
	Moreover, the same description is the basis for our numerical toolbox, including exact expressions for, among others, expectation values, state fidelities, heralding success probabilities, and even nonclassical quantum correlations.
	Our technique applies equally to state-based and measurement-based quantum photonics.

	Furthermore, our method naturally accounts for many imperfections.
	For instance, our framework enables us to directly and exactly include, describe, and quantify finite quantum efficiencies, noise count contributions, saturation effects in detection schemes with finite photon-number resolutions, higher-order photon-number contributions of heralded photon-number states, excess noise from amplification, etc.
	Since we have access to arbitrary combinations of such imperfections, we can, in turn, use our method to propose improvements to mitigate the negative impact of such perturbations.

	For instance, we found that heralding with an improved photon-number resolution is advantageous even if we condition to single clicks only, and that amplifiers can be useful even in the few-photon regime when properly balancing excess noise against gain rather than choosing a gain which fully counteracts losses.
	Both improvements are to some extend counterintuitive in a classical picture.
	Within our full quantum description, however, it certainly makes sense that measurements have a projective effect onto states even if only parts of a measurement are used, and that an increased gain also leads to additional and unavoidable noise contributions.
	We exemplified our general finding with these specific examples, modeling realistic experiments together with readily available sources of quantum light.
	However, we emphasize that our approach is not limited to these scenarios and can be applied to other cases as well.

	As such, our method is not only the theoretical basis for future experimental implementations, it also serves as a starting point for future theoretical studies;
	we briefly outlined a few of them.
	While we considered a first example in which we quantified quantum correlations of a two-mode light field, a full multimode description could be developed to generalize the results found here.
	For instance, correlated losses and nonlinear processes which are not mode-matched with a signal field are of additional interest for many experiments.
	Similarly, other quantum-optical nonlinearities could be studied, e.g., to analyze a Kerr-type medium.
	Also, we mainly focused on nonclassical effects as defined in quantum optics.
	Other types of quantum phenomena and their optimization in photonic systems, such as multiphoton entanglement, are of major relevance for future quantum technologies, such as quantum communication in large optical networks and could be accessible with our method.

	In conclusion, a framework has been devised for the description for photons traveling in networks which include feedback loops and active elements.
	Our method is well suited for the realistic description and directed planning of experimental setups, aiming at advancing their performance by maximizing quantum features while also minimizing experimental resources and constraints.
	Furthermore, our techniques could inspire future extensions, which hopefully further advance the realization of practical quantum technologies and mark the starting point for exploring the full potential of quantum photonics with active feedback loops.

	\quad
\begin{acknowledgments}
	The Integrated Quantum Optics group acknowledges financial support through the Gottfried Wilhelm Leibniz-Preis (Grant No. SI1115/3-1) and the European Commission through the ERC project \mbox{QuPoPCoRN} (Grant No. 725366).
	M. B. P. and I. D. acknowledge support through the ERC Synergy grant \mbox{BioQ} (Grant No. 319130).
\end{acknowledgments}

\appendix

\section{Commutation rules and partial trace}\label{app:OpAlg}

	We consider an operator reordering for exponential operators.
	The applied technique is based on the single-mode approach in Ref. \cite{SW18}, therein Appendix A.
	With that method, we can easily verify
	\begin{equation}
	\begin{split}
		\exp\left(x\hat a\otimes\hat a\right)u^{\hat n}\otimes v^{\hat n}
		=&
		u^{\hat n}\otimes v^{\hat n}\exp\left(xuv\hat a\otimes\hat a\right),
		\\
		u^{\hat n}\otimes v^{\hat n} \exp\left(y\hat a^\dag\otimes\hat a^\dag\right)
		=&
		\exp\left(yuv\hat a^\dag\otimes\hat a^\dag\right) u^{\hat n}\otimes v^{\hat n},
	\end{split}
	\end{equation}
	for $u,v,x,y\in\mathbb C$.

	In addition, we require a reordering of exponential functions of $\hat a^\dag\otimes\hat a^\dag$ and $\hat a\otimes\hat a$.
	For the following calculations, it is relevant to recall the following simple relations:
	a decomposition of the identity in terms of coherent states, $\pi\hat 1=\int_\mathbb C d^2\alpha |\alpha\rangle\langle\alpha|$;
	a normally ordered representation of coherent states, ${:}\exp([\hat a-\alpha]^\dag[\hat a-\alpha]){:}=|\alpha\rangle\langle\alpha|$;
	a special case of the Baker-Campbell-Hausdorff formula, $\exp(u\hat a)\exp(v\hat a^\dag)=\exp(uv)\exp(v\hat a^\dag)\exp(u\hat a)$;
	and a Gaussian integral identity, $\int_{\mathbb C} d^2\alpha \exp(-w|\alpha|^2+u\alpha^\ast+v\alpha)=(\pi/w)\exp(uv/w)$ for $\mathrm{Re}(w)>0$.
	It is also worth emphasizing that, under normal ordering, operators behave like complex numbers \cite{VW06}.
	Applying the above relations for $1>\mathrm{Re}(xy)$ results in
\begin{widetext}
	\begin{equation}
	\begin{split}
		\exp\left(x\hat a\otimes\hat a\right)\exp\left(y\hat a^\dag\otimes\hat a^\dag\right)
		=& \int_{\mathbb C} \frac{d^2\alpha}{\pi} \exp\left(x\hat a\otimes\hat a\right)|\alpha\rangle\langle\alpha|\otimes\hat 1\exp\left(y\hat a^\dag\otimes\hat a^\dag\right)
		\\
		=& \int_{\mathbb C} \frac{d^2\alpha}{\pi} {:}\exp([\hat a{-}\alpha]^\dag[\hat a{-}\alpha]){:}\otimes\exp\left(x\alpha\hat a\right)\exp\left(y\alpha^\ast\hat a^\dag\right)
		\\
		=& {:}\frac{e^{-\hat n\otimes \hat 1}}{\pi} \int_{\mathbb C} d^2\alpha \exp\left(
			-[1-xy]|\alpha|^2
			+[\hat a\otimes\hat 1+y\hat 1\otimes\hat a^\dag]\alpha^\ast
			+[\hat a^\dag\otimes\hat 1+x\hat 1\otimes\hat a]\alpha
		\right){:}
		\\
		=& \frac{1}{1-xy}{:}\exp\left(
			\frac{y\hat a^\dag\otimes\hat a^\dag}{1-xy}
			+\frac{xy\left[\hat n\otimes\hat 1+\hat 1\otimes\hat n\right]}{1-xy}
			+\frac{x\hat a\otimes\hat a}{1-xy}
		\right){:}
		\\
		=&
		\exp\left(\frac{y\hat a^\dag\otimes\hat a^\dag}{1-xy}\right)
		\left(\frac{1}{1-xy}\right)^{\hat n\otimes\hat 1+\hat 1\otimes\hat n+\hat 1\otimes\hat 1}
		\exp\left(\frac{x\hat a\otimes\hat a}{1-xy}\right).
	\end{split}
	\end{equation}
\end{widetext}

	Finally, we may compute the partial trace of the operators considered so far.
	Applying the same techniques as used above, we find
	\begin{equation}
	\begin{split}
		& \mathrm{id}\otimes\mathrm{tr}\left[
			e^{y\hat a^\dag\otimes\hat a^\dag}
			u^{\hat n}\otimes v^{\hat n}
			e^{x\hat a^\dag\otimes\hat a^\dag}
		\right]
		\\
		=& \int_{\mathbb C} \frac{d^2\alpha}{\pi}
		\left(\hat 1\otimes\langle\alpha|\right)
			e^{y\hat a^\dag\otimes\hat a^\dag}
			u^{\hat n}\otimes v^{\hat n}
			e^{x\hat a^\dag\otimes\hat a^\dag}
		\left(\hat 1\otimes|\alpha\rangle\right)
		\\
		=&\frac{1}{1-v}
		\left(u+\frac{xy}{1-v}\right)^{\hat n}.
	\end{split}
	\end{equation}

\section{Noise counts and click-counting moments}\label{app:Clicks}

	In Eq. \eqref{eq:OnOff}, the impact of the noise contribution on a single on-off detector is shown.
	More generally, a measurement operator for $K'$ clicks from a multiplexing of $N$ on-off detectors with a noise count contribution $\delta$ is given by
	\begin{equation}
		\hat\Pi_{K'}^{(\eta,\delta)}=\sum_{K=K'}^N\binom{K}{K'}\left(e^{-\delta}\right)^{K'}\left(1-e^{-\delta}\right)^{K-K'}\hat\Pi_{K}^{(\eta,0)},
	\end{equation}
	where $\hat\Pi_{K}^{(\eta,0)}$ labels the analog noise-free operator for $K$ clicks, which includes a quantum efficiency $\eta$.
	It is further worth noting that applying the same type of convolution to $\hat\Pi_{K'}^{(\eta,\delta)}$ with a negative count rate, $-\delta$, allows one to deconvolute dark counts to retrieve $\hat\Pi_{K}^{(\eta,0)}$ \cite{LSV15}.
	Moreover, $\hat\Pi_{K'}^{(\eta,\delta)}$ is a linear combination of operators $\hat E(z)$ since $\hat\Pi_{K}^{(\eta,0)}$ is.

	In addition to the treatment of dark counts, we may also briefly summarize moment-based nonclassicality criteria for click-counting detectors which have been rigorously derived in Ref. \cite{SVA13}.
	In particular, we consider second-order criteria.
	For instance, the following variance-based constraint holds true for classical states:
	\begin{equation}
	\begin{split}
		& \mathrm{tr}[\hat\rho{:}(\Delta e^{[w-1]\hat n})^2{:}]
		\\
		=&\mathrm{tr}[\hat\rho{:}(e^{[w-1]\hat n})^2{:}]-\left(\mathrm{tr}[\hat\rho{:}e^{[w-1]\hat n}{:}]\right)^2
		\\
		=& \det\begin{pmatrix}
			\mathrm{tr}[\hat\rho\hat E(1)] & \mathrm{tr}[\hat\rho\hat E(w)]
			\\
			\mathrm{tr}[\hat\rho\hat E(w)] & \mathrm{tr}[\hat\rho\hat E(2w-1)]
		\end{pmatrix}
		\geq0,
	\end{split}
	\end{equation}
	where we applied the exponential measurement operators $\hat E$ as used throughout this paper.
	A violation of this inequality certifies nonclassical light, termed sub-binomial light \cite{SVA12}.
	Note that we choose $w=1/2$ in the main text for simplicity.
	Analogously, including the chosen setting $z=1/2$, a cross-correlation-based constraints for classical states can be formulated \cite{SVA12} and applied,
\begin{widetext}
	\begin{equation}
	\begin{split}
		&
		\left(\mathrm{tr}\otimes\mathrm{tr}[\hat\rho{:}(\Delta e^{[w-1]\hat n})^2{:}\otimes\hat 1]\right)
		\left(\mathrm{tr}\otimes\mathrm{tr}[\hat\rho\hat 1\otimes{:}(\Delta e^{[z-1]\hat n})^2{:}]\right)
		-\left(
			\mathrm{tr}\otimes\mathrm{tr}[\hat\rho{:}\Delta e^{[w-1]\hat n}{:}\otimes{:}\Delta e^{[z-1]\hat n}{:}]
		\right)^2
		\\
		= &
		\det\begin{pmatrix}
			\mathrm{tr}\otimes\mathrm{tr}[\hat\rho\hat E(1)\otimes\hat E(1)]
			& \mathrm{tr}\otimes\mathrm{tr}[\hat\rho\hat E(w)\otimes\hat E(1)]
			& \mathrm{tr}\otimes\mathrm{tr}[\hat\rho\hat E(1)\otimes\hat E(z)]
			\\
			\mathrm{tr}\otimes\mathrm{tr}[\hat\rho\hat E(w)\otimes\hat E(1)]
			& \mathrm{tr}\otimes\mathrm{tr}[\hat\rho\hat E(2w-1)\otimes\hat E(1)]
			& \mathrm{tr}\otimes\mathrm{tr}[\hat\rho\hat E(w)\otimes\hat E(z)]
			\\
			\mathrm{tr}\otimes\mathrm{tr}[\hat\rho\hat E(1)\otimes\hat E(z)]
			& \mathrm{tr}\otimes\mathrm{tr}[\hat\rho\hat E(w)\otimes\hat E(z)]
			& \mathrm{tr}\otimes\mathrm{tr}[\hat\rho\hat E(1)\otimes\hat E(2z-1)]
		\end{pmatrix}
		\geq 0.
	\end{split}
	\end{equation}
\end{widetext}


\end{document}